\documentclass[10pt,aps,prd,twocolumn,a4paper,showpacs,superscriptaddress,nofootinbib]{revtex4-1}
\usepackage{amsmath}
\usepackage{amsfonts}
\usepackage[colorlinks=true,linkcolor=blue,citecolor=blue,urlcolor=blue]{hyperref}
\usepackage{amssymb}
\usepackage{graphicx}
\usepackage{csquotes}
\usepackage{color}

\renewcommand{\Re}{\textrm{Re}}
\renewcommand{\Im}{\textrm{Im}}
\newcommand{\bk}{\mathbf{k}}
\newcommand{\eq}[1]{Eq.~\eqref{#1}}
\providecommand{\abs}[1]{\left|#1\right|}
\providecommand{\ep}[1]{{e}^{#1}}
\newcommand{\xx}{\mathbf{x}}

\begin{document}
\title{Spectrum and entanglement of phonons in quantum fluids of light} 

\author{Xavier Busch}
\email[]{xavier.busch@th.u-psud.fr}
\affiliation{Laboratoire de Physique Th\'eorique, CNRS UMR 8627, B{\^{a}}timent\ 210,
         \\Universit\'e Paris-Sud 11, 91405 Orsay CEDEX, France.}

\author{Iacopo Carusotto}\email[]{carusott@science.unitn.it}
\affiliation{INO-CNR BEC Center and Dipartimento di Fisica, Universit\`a di Trento, via Sommarive 14, 38123 Povo, Italy.}

\author{Renaud Parentani}
\email[]{renaud.parentani@th.u-psud.fr}
\affiliation{Laboratoire de Physique Th\'eorique, CNRS UMR 8627, B{\^{a}}timent\ 210,
         \\Universit\'e Paris-Sud 11, 91405 Orsay CEDEX, France.}
         

\begin{abstract}
We study the quantum state of phonons propagating on top of a fluid of light coherently generated in a planar microcavity device by a quasi-resonant incident laser beam. In the steady-state under a monochromatic pump, because of the finite radiative lifetime of photons, a sizable incoherent population of low-frequency phonons is predicted to appear. Their mean occupation number differs from a Planck distribution and is independent of the photon lifetime. When the photon fluid is subjected to a sudden change of its parameters, additional phonon pairs are created in the fluid with remarkable two-mode squeezing and entanglement properties. Schemes to assess the nonseparability of the phonon state from measurements of the correlation functions of the emitted light are discussed. 
\end{abstract}

\maketitle

\section{Introduction}

Among the rich features of quantum fluids of atoms~\cite{pitaevskii2003bose,trento_fermi_RMP,bloch_dalibard_RMP} and of light~\cite{Carusotto:2012vz} a remarkable position is held by the intriguing structure of their vacuum state and by the possibility of accurately measuring its static and dynamical properties under various conditions. In particular, when such a fluid is subjected to some temporal and/or spatial change of its properties, zero-point fluctuations of its quantum vacuum state are converted into correlated pairs of propagating quasi-particles. As pointed out in Ref.~\cite{Unruh:1980cg}, this offers the possibility of conceiving experiments aiming to test longstanding predictions concerning quantum processes related to cosmology and to black holes. For instance, one may consider observing the emitted pairs associated with the (analogous) cosmological pair creation or dynamical Casimir (DCE) effects~\cite{Campo:2003pa,Fedichev:2003bv,Carusotto:2009re}, as well as those produced by the (analogous) Hawking effect~\cite{Massar:1995im,Brout:1995wp,Balbinot:2007de,Carusotto:2008ep,Macher:2009nz,Recati:2009ya,Gerace:2012an}. Most remarkably, accurate measurements of the observables are expected to distinguish quantum (spontaneous) from classical (stimulated) correlations by looking at the non-separability of the outgoing state~\cite{Campo:2005sy,Horstmann:2010xd,Adamek:2013vw,Lahteenmaki12022013,Bruschi:2013tza,Busch:2013sma,finazzi2013}, or violations of Cauchy-Schwarz~\cite{PhysRevLett.108.260401,deNova:2012hm} or even Bell inequalities~\cite{Campo:2005sv}.

In recent years, many material platforms have been proposed and investigated in this context, in particular ultracold atomic gases and superfluid liquid helium~\cite{Barcelo:2005fc}. Very recently, quantum fluids of light in planar microcavity devices have been recognized as most promising candidates for experimental studies of the quantum vacuum~\cite{marino,sarchi,fleurov,solnyshkov,Gerace:2012an}. In one- or two-dimensional microcavity devices, photons acquire an effective mass $m$ because of spatial confinement, while an effective two-body photon-photon interaction can originate from the $\chi^{(3)}$ optical nonlinearity of the cavity medium. As a result, assemblies of many photons in the cavity can display the collective behavior of a Bose-Einstein condensate with a macroscopic occupation of a single quantum state and a long-range coherence, as well as superfluid hydrodynamic features with the low-frequency elementary excitations having a collective phonon nature. In order to reinforce the interactions, one often works with microcavities in the so-called strong light-matter coupling regime, where the photon mode is coupled to a resonant material excitation, typically of excitonic nature. For the purpose of the present article, the resulting {\em polariton} excitations can be simply understood as dressed photons with an enhanced nonlinearity. 

In contrast to material systems where the lifetime of the basic constituents is virtually infinite, cavity photons are intrinsically subject to losses due, e.g., to the imperfect reflectivity of the cavity mirrors. On one hand, these losses are experimentally very useful as they allow continuous measurement in a non destructive manner of the photon state. On the other hand, they introduce new features, for instance the phonons can acquire a finite mass and the photon fluid requires a continuous external pumping to compensate losses: among the different available schemes, here we shall restrict our attention to the case of a coherent pump quasi-resonant with the cavity mode.

The present work reports a theoretical investigation of the quantum fluctuations of the quantum fluid of light and, in particular, of its phonon excitations. Even though the same results can be derived within a Wigner formalism~\cite{Carusotto:2008ep,Gerace:2012an,plimak,Koghee2013}, our presentation here will be based on a quantum Langevin description~\cite{Caldeira:1981rx,Unruh:1989dd,gardiner2004quantum,PhysRevA.74.033811,Parentani:2007uq} of the driven-dissipative photon fluid because it directly provides reliable predictions for the separability of the system state~\cite{Adamek:2013vw,Busch:2013sma}. In the simplest geometry with a spatially homogeneous background solution, explicit expressions for the correlation functions can be obtained by fully analytical means for both the stationary state and the temporal response to a sudden jump in the system parameters. 

The article is organized as follows: In Sec.~\ref{sec:model} we present the physical system under consideration and we quickly review the equations of motion ruling the quantum fluid of light. The stationary state is then studied in Sec.~\ref{sec:equilibrium}. Inspired by the analogy to cosmological pair creation and dynamical Casimir effects, the response of the system to a sudden change of its parameters is discussed in Sec.~\ref{sec:DCE}. Conclusions are drawn in Sec.~\ref{sec:Conclusion}. 

\section{The physical system and the model}
\label{sec:model}

\begin{figure}
\includegraphics[width=0.9\linewidth]{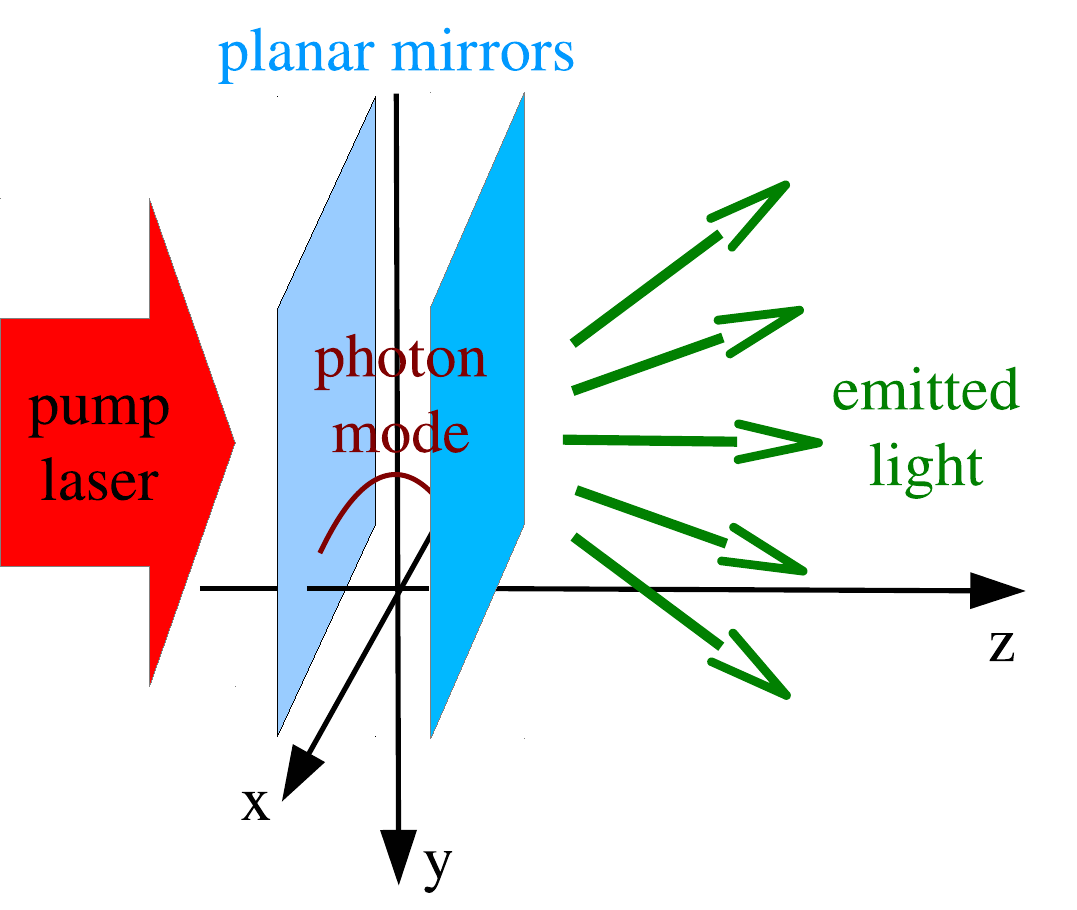}
\caption{(Color online) Sketch of the planar microcavity system under consideration.}
\label{fig:sketch}
\end{figure}

\subsection{The Hamiltonian of the system}
A sketch of the planar microcavity system we are considering is shown in Fig.~\ref{fig:sketch}. A comprehensive review of its rich physics can be found in~\cite{Carusotto:2012vz}; here we briefly summarize the main features that are important for our discussion. In the simplest configuration, light is confined in a cavity material of refractive index $n_0$ sandwiched between two high-quality plane-parallel metallic mirrors spaced by a distance $\ell_z$. Photon propagation along the $z$-axis is then quantized as $q_z=\pi M/ \ell_z$, $M$ being a positive integer. For each longitudinal mode $M$, the frequency dispersion of the mode as a function of the in-plane wave vector $\bk$ has the form
\begin{equation}
E_{\rm cav}(k) = \frac{\hbar c}{n_0} \sqrt{q_z^2 + k^2}\simeq E_0^{\rm bare} + \frac{\hbar^2 k^2}{2 m},
\label{eq:omega_cav}
\end{equation}
where the effective mass $m$ of the photon and the rest energy $E_0^{\rm bare}$ are related by the relativistic-like expression
\begin{equation}
m= \frac{\hbar q_z}{c/ n_0}= \frac{E_0^{\rm bare}}{c^2/n_0^2}.
\label{eq:m_cav}
\end{equation}
Neglecting for simplicity the polarization degrees of freedom, we can define the creation and destruction operators $\hat a^\dagger_{\bk}$ and $\hat a_{\bk}$ for each mode of wavevector $\bk$ and their real-space counterparts 
\begin{equation}
\label{eq:phi}
 \hat \Phi(\xx)= \int \frac{d\bk}{(2\pi)^{d/2}} \ep{i \bk \xx}\, a_{\bk},
\end{equation}
which satisfy the usual equal-time commutation (ETC) rules $[\hat \Phi(\xx), \hat \Phi^\dagger(\xx')] = \delta(\xx-\xx')$ of a non-relativistic quantum field. In Eq.~\eqref{eq:phi}, $d$ is the dimensionality of the fluid along the cavity: while standard planar cavities such as the one sketched in Fig.~\ref{fig:sketch} have $d=2$, effective one-dimensional $d=1$ fluids can be created with an additional in-plane confinement~\cite{Carusotto:2012vz}. In term of the quantum field operator $\hat \Phi(\xx)$, the isolated cavity Hamiltonian in units where $\hbar=1$ can be written in the form
\begin{equation}
\label{eq:hamiltonian0}
H_0\! = \! \! \int \! \! d\xx  \left [  E_0^{\rm bare} \hat \Phi^\dagger \hat \Phi+ \frac{1}{2m} (\nabla_\xx\hat \Phi^\dagger) (\nabla_\xx\hat \Phi)  + \frac{g}{2} \hat \Phi^\dagger \hat \Phi^\dagger \hat \Phi \hat \Phi  \right].
\end{equation}
The first two terms describe the photon rest energy and its effective (kinetic) mass, respectively. The last term accounts for a Kerr optical nonlinearity of the cavity medium which is essential to have sizable photon-photon interactions. The $g$ coefficient quantifying the interaction strength is proportional to the material $\chi^{(3)}$: explicit expressions can be found in the quoted review article.

In addition to its conservative internal dynamics ruled by $H_0$, the cavity is coupled to external baths including, e.g., the radiative coupling to the propagating photon modes outside the cavity via the (small) transmittivity of the mirrors. A typical way of modeling the dissipative effects due to this environment is based on a Hamiltonian formalism where the environment is described by a phenomenological quantum field $ \hat{ \Psi}_\zeta$. The bath and interaction Hamiltonians have the forms
\begin{subequations}
\label{eq:hamiltonian_bath}
\begin{align}
H_{\rm bath} \!&= \! \! \int \! \! d\xx  \int\! \!  d\zeta\, \omega_\zeta \hat{ \Psi}_\zeta^\dagger(\xx)  \hat{ \Psi}_\zeta(\xx)  \\ 
H_{\rm int} \!&= \! \! \int \! \! d\xx  \int\! \!  d\zeta\, \left [
g_\zeta \left ( \hat \Phi^\dagger(\xx)\, \hat {\Psi}_\zeta(\xx)  + \hat \Phi(\xx) \, \hat {\Psi}_\zeta ^\dagger(\xx) \right )\right ].
\end{align}
\end{subequations}
Here, the bath operators $ \hat{ \Psi}_\zeta$ obey the usual ETC $[\hat{ \Psi}_\zeta(\xx,t),\hat{ \Psi}_{\zeta'}^\dagger (\xx',t)] = \delta(\xx-\xx')\, \delta(\zeta - \zeta')$ and the index $\zeta$ is a continuous wave number: this ensures that the $\hat \Psi$ field forms a dense set of degrees of freedom, a condition necessary to obtain dissipation in Hamiltonian systems~\cite{Caldeira:1981rx,Unruh:1989dd,gardiner2004quantum}. In the case of radiative loss processes from a planar microcavity, the $\hat{ \Psi}_\zeta(\xx,t)$ operator corresponds to the destruction operator of extra-cavity photons and the $\zeta$ quantum number indicates the value of the normal component of the extra-cavity wave vector. While more realistic descriptions of the microcavity device can be used to obtain first-principle predictions for the $g_\zeta$ coupling constant~\cite{PhysRevA.74.033811}, in the present work we consider it as a (real-valued) model parameter to be adjusted so as to reproduce the experimentally observed photon decay rate $\Gamma$; see below \eq{Gamm}. In the quantum optical literature, approaches to dissipation based on Hamiltonians of the form (\ref{eq:hamiltonian_bath}) go under the name of input-output formalism~\cite{PhysRevA.74.033811,d1995quantum,gardiner2004quantum}. With respect to equivalent descriptions based on the truncated Wigner distribution~\cite{Gerace:2012an,Koghee2013}, this method has the main advantage that unitarity is manifestly preserved as $\hat \Phi$ and $ \hat{ \Psi}_\zeta$ are treated on an equal footing.

As a last step, we have to include the Hamiltonian term describing the coherent pumping of the cavity by an incident laser field of (normalized) amplitude $F(\xx,t)$,
\begin{equation}
\label{eq:hamiltonian_F}
H_F\! = \! \! \int \! \! d\xx\,\left[\hat \Phi(\xx)\, F^*(\xx,t) + \hat \Phi^\dagger(\xx)\, F(\xx,t)\right].
 \end{equation}
For a monochromatic pump of frequency $\omega_P$, wavevector $\bk_P$, and a very wide waist, we can perform a plane wave approximation and write
\begin{equation}
 F(\xx,t)=F_0\,e^{-i\omega_P t}\,e^{i\bk\cdot\xx},
\end{equation}
where $\bk$ is the projection of $\bk_P$ along the cavity plane. In the following, we shall restrict our attention to the case of a monochromatic pump normally incident on the cavity, which gives a vanishing in-plane $\bk=0$ and therefore a spatially homogeneous and isotropic pump amplitude $F(\xx,t)=F_0(t)\,e^{-i\omega_p t}$: this pump configuration injects into the cavity a photon fluid that is spatially homogeneous and at rest. Even though the coherent pump acts on the single $\bk=0$ mode, the presence of the interaction term causes the field dynamics to involve the whole continuum of in-plane $\bk$ modes.

To summarize, the total Hamiltonian has the form
\begin{equation}
\label{eq:hamiltonian}
\begin{split}
H\! = \! \! &\int \! \! d\xx  \left [  E_0^{\rm bare} \hat \Phi^\dagger \hat \Phi+ \frac{1}{2m} (\nabla_\xx\hat \Phi^\dagger) (\nabla_\xx\hat \Phi) + \frac{g}{2} \hat \Phi^\dagger \hat \Phi^\dagger \hat \Phi \hat \Phi  \right. \\
& + \left. \hat \Phi F^* + \hat \Phi^\dagger F + \!\! \int\! \!  d\zeta \omega_\zeta \hat{ \Psi}_\zeta^\dagger  \hat{ \Psi}_\zeta + g_\zeta \left ( \hat \Phi^\dagger \hat {\Psi}_\zeta  + \hat \Phi \hat {\Psi}_\zeta ^\dagger \right )\right ].
\end{split}
\end{equation}

\subsection{The equations of motion}

From Eq.~\eqref{eq:hamiltonian}, the equations of motion are 
\begin{subequations}
\label{eq:eombrut}
\begin{align}
\label{eq:eompsi}
i\partial_t \hat \Phi  & = \left ( E_0^{\rm bare} - \frac{\nabla_\xx^2}{2m} + g \hat \Phi^\dagger \hat \Phi\right ) \hat \Phi +  \int d\zeta\, g_\zeta  \hat {\Psi}_\zeta  +F , \\
\label{eq:eomX}
i \partial_t \hat {\Psi}_\zeta& =  \omega_\zeta  \hat {\Psi}_\zeta+  g_\zeta \hat \Phi  . 
\end{align}
\end{subequations}
The solution of Eq.~\eqref{eq:eomX} can be written as
\begin{equation}
\label{eq:solX}
 \begin{split}
\hat {\Psi}_\zeta(\xx,t) &= \hat {\Psi}_\zeta^0(\xx,t) -i \int dt' \theta(t-t') \ep{ - i\omega_\zeta (t-t')} g_\zeta \hat \Phi (\xx,t').
 \end{split}
 \end{equation} 
The first term is the homogeneous solution
\begin{equation}
 \hat {\Psi}_\zeta^0 (\xx,t) =\hat c(\xx, \zeta) \, \ep{- i \omega_\zeta t}. 
\end{equation}
Here $\hat  c(\xx,\zeta)$ is the destruction operator of a environment quantum of energy $\omega_\zeta$ localized at $\xx$. It obeys the canonical commutator 
\begin{equation}
[ \hat  c(\xx',\zeta'), \hat c^\dagger(\xx,\zeta) ] =\delta(\xx-\xx')\, \delta(\zeta- \zeta') . 
\end{equation}
 
Introducing the right hand side  of \eq{eq:solX} in Eq.~\eqref{eq:eompsi} gives the effective equation of motion for the photon field,
\begin{equation}
\label{eq:eomdissip}
\begin{split}
i\partial_t \hat \Phi &= \left ( E_0^{\rm bare} - \frac{\partial_x^2}{2m} + g \hat \Phi^\dagger \hat \Phi\right) \hat \Phi \\
&- i  \int dt'  D(t-t') \hat \Phi (t')  +  \int d\zeta g_\zeta  \hat {\Psi}_\zeta^0 + F =0 .
\end{split}
\end{equation} 
The non local dissipative kernel is
\begin{equation}
\label{eq:disskernel}
\begin{split}
 D(t-t') &\doteq \theta(t-t') \int d\zeta g_\zeta^2\ep{ - i\omega_\zeta (t-t')} ,
\end{split}
\end{equation} 
and its Fourier transform is 
\begin{equation}
\begin{split}
 \tilde D(\omega) &= \int d\zeta g_\zeta^2 \frac{i}{\omega - \omega_\zeta + i \epsilon} . 
\label{tDom}
\end{split}
\end{equation} 
Because of the high frequency of the pump as compared to the time scale of the hydrodynamic evolution of the fluid, we shall see below that $D(t-t')$ can be well approximated by a local kernel within a sort of {\em Markov} approximation and correspondingly $\tilde D(\omega)$ can be approximated by a constant value independent of $\omega$.

Under the weak-interaction assumption, we perform the usual dilute gas approximation~\cite{pitaevskii2003bose} and we split the field operator as the sum $\hat \Phi = \bar \Phi + \delta\hat\Phi$ of a (large) coherent component $\bar \Phi(\xx,t)$ corresponding to the condensate and a (small) quantum fluctuation field $\delta \hat \Phi(\xx,t)$. Including the new terms stemming from pumping and from losses, the mean field  $\bar \Phi(\xx,t)$ can be shown to obey a generalized Gross-Pitaevskii-Langevin equation of the form
\begin{equation}
\label{eq:GPEgeneral}
\begin{split}
i\partial_t \bar \Phi(\xx,t) M= &\left (  E_0^{\rm bare} - \frac{\partial_x^2}{2m} + g \abs{\bar \Phi}^2\right ) \bar \Phi(\xx,t) \\
&- i  \int dt'  D(t-t') \bar \Phi (\xx,t') + F(\xx,t).
\end{split}
\end{equation}
When assuming that $F(\xx,t)=F_0(\xx,t) \, e^{- i \omega_p t}$ and  $\bar \Phi(\xx,t) =\bar \Phi_0(\xx,t) \, e^{- i \omega_p t}$ with $\bar \Phi_0(\xx,t)$ and $F_0(\xx,t)$ slowly varying functions of time, it is appropriate to extract the temporally local part of the dissipative kernel and to rewrite Eq.~\eqref{eq:GPEgeneral} as 
\begin{multline}
\label{eq:GPEomegae}
i\partial_t \bar \Phi_0(\xx,t) = F_0(\xx,t) + \\ 
\quad \left ( E_0^{\rm bare} + \Delta E - \omega_p - \frac{\partial_x^2}{2m} + g \abs{\bar \Phi_0}^2 - i\Gamma \right) \bar \Phi_0(\xx,t)\\  
\quad - i \int dt'  D(t-t') e^{i \omega_p (t-t')} \left [\bar \Phi_0(\xx,t')  - \bar \Phi_0(\xx,t) \right ]  ,
\end{multline}
where the real and imaginary parts of $\tilde D( \omega_p)=\Gamma+i\Delta E$ defined in \eq{tDom} respectively give the decay rate $\Gamma$ and a (small) shift $\Delta E $ of the photon frequency; see Ref.~\cite{Caldeira:1981rx}. Explicitly, one has 
\begin{equation}
\label{Gamm}
\begin{split}
\Delta E &= -  \int d\zeta g_\zeta^2 P.V.\frac{1}{ (\omega_\zeta- \omega_{\rm p}) } ,\\
 \Gamma &= \int d\zeta \, g_\zeta^2 \, \pi \delta(\omega_\zeta- \omega_{\rm p} ) ,
\end{split}
\end{equation} 
where $P.V.$ is the principal value. In the following, all formulas will be written in terms of the effective cavity photon frequency, $E_0=E_0^{\rm bare}+\Delta E$. The ratio $ \omega_p /\Gamma =Q_p$ gives the quality factor of the cavity: in typical microcavity systems one has $Q_p \lesssim 10^5$. For a given $\omega_p$, various choices of $g_\zeta$ giving the same value for $\Gamma$ should be considered at this level as physically equivalent. 

\section{The stationary state} 
\label{sec:equilibrium}

We begin our discussion of quantum fluctuations in a stationary state under a spatially homogeneous and monochromatic pump at frequency $\omega_p$, $F(\xx,t)=F_0\, e^{- i \omega_p t}$ with a constant pump amplitude $F_0$. In this case, we can safely assume the coherent component $\bar \Phi$ of the photon field to be itself spatially homogeneous and monochromatically oscillating at $\omega_p$, $\bar \Phi(\xx,t) = {\mathtt \Phi}_0\, \ep{ - i \omega_p t}$. Using Eq.~\eqref{eq:GPEomegae}, it is immediate to see that ${\mathtt \Phi}_0$ obeys the state equation
\begin{equation}
\label{eq:GPEhomo}
\begin{split}
\left [ \omega_p - E_0 - g \abs{{\mathtt \Phi}_0}^2 + i\Gamma \right ] {\mathtt \Phi}_0 =  F_0.
\end{split}
\end{equation}
In the following, we shall assume that the phase of the pump $F_0$ is chosen in such a way as to give a real and positive ${\mathtt \Phi}_0>0$. As we are interested in a stable configuration where the phonon mass is the smallest, we will follow previous work on analogous models based on superfluids of light in microcavities~\cite{Gerace:2012an} and concentrate our attention on the case of a pump frequency blue detuned with respect to the bare photon frequency $\omega_p>E_0$, where the dependence of the internal intensity $|\mathtt{\Phi}_0|^2$ on the pump intensity $|F_0|^2$ shows a bistability loop~\cite{PhysRevLett.93.166401,Carusotto:2012vz}. More specifically, we shall concentrate on the upper branch of the bistability loop, where interactions have shifted the effective photon frequency $E_0+g|\mathtt{\Phi}_0|^2$ to the blue side of the pump laser, $E_0+g|\mathtt{\Phi}_0|^2 \geq \omega_p$. Exact resonance $\omega_p=E_0+g|\mathtt{\Phi}_0|^2$ is found at the end point of the upper branch of the bistability loop: as we shall see shortly, only this point corresponds to a vanishing phonon mass. The more complex physics of quantum fluctuations under a monochromatic pump in the vicinity of the so-called ``magic angle'' was discussed in~\cite{sarchi} for pump intensities spanning across the optical parametric oscillation threshold~\cite{Ciuti:SST2003}.

\subsection{The equation of motion} 

Equations~\eqref{eq:eomdissip} and~\eqref{eq:GPEgeneral} determine the equation for linear perturbations $\hat \delta \Phi$. Taking into account the spatial homogeneity of the mean-field solution $\bar \Phi$, we use the relative perturbation $\hat \phi_\bk = \hat{\delta\Phi}_\bk /\mathtt{\Phi}_0$ at given wave number $\bk$. Using a Markovian\footnote{
The exact equation is given in Appendix~\ref{app:FD} as Eq.~\eqref{eq:perturb}. Since the characteristic frequency $\omega_k$ of phonon modes is much lower than $\omega_p$, the non-local part of the dissipative term of Eq.~\eqref{eq:disskernel} can be neglected as it gives corrections proportional to $\omega_k/\omega_p$. In fact, using Eq.~\eqref{tDom}, $[\tilde D(\omega_p + \omega) -  \tilde D(\omega_p)] \sim \tilde D(\omega_p)\, \omega/ \omega_p$ for typical dissipation baths, which is much smaller in magnitude than $\tilde D(\omega_p)$ when $\omega\ll \omega_p$.} 
approximation to neglect the non-local part of the dissipative term, one obtains the following quantum Langevin equation of motion:
\begin{equation}
\label{eq:2x2eom}
\begin{split}
i(\partial_t  + \Gamma  )\hat \phi_\bk &= \Omega_k \hat \phi_\bk+  m c^2 \hat \phi_{-\bk}^\dagger +
\frac{\hat S_\bk}{\abs{{\mathtt \Phi}_0}}.
\end{split}
\end{equation}
Its conservative part shows interesting differences from the case of atomic condensates: While the interaction energy has the same form
\begin{equation}
 mc^2\doteq g |\mathtt{\Phi}_0|^2,
\label{inten}
\end{equation}
the detuning coefficient multiplying $\hat \phi_\bk$ in Eq.~\eqref{eq:2x2eom} keeps track of the pump frequency $\omega_p$. It is given by
\begin{equation}
\Omega_k \doteq \frac{k^2}{2m} - \omega_p + E_0 + 2 m c^2 ,
\label{eq:defOmega}
\end{equation}
and allows for a larger variety of Bogoliubov dispersions~\cite{PhysRevLett.93.166401,Carusotto:2012vz}. The eigenmodes of the deterministic part of the linear problem described by \eq{eq:2x2eom} are in fact characterized by the dispersion
\begin{equation}
\label{eq:omk}
\begin{split}
\omega_k^2 \doteq \Omega_k^2-m^2c^4 , 
\end{split}
\end{equation}
and a collective phonon destruction operator of the form
\begin{equation}
\label{eq:bogotf}
\begin{split}
\hat \varphi_\bk &\doteq {\mathtt \Phi}_0 \left ( u_k \hat \phi_\bk + v_k \hat \phi_{-\bk}^\dagger \right ) ,\\ 
u_k &\doteq  \frac{\sqrt{\Omega_k+mc^2}+ \sqrt{\Omega_k-mc^2}}{2\sqrt{\omega_k}} \\ 
v_k &\doteq  \frac{\sqrt{\Omega_k+mc^2}- \sqrt{\Omega_k-mc^2}}{2\sqrt{\omega_k}}.
\end{split}
\end{equation} 
Using Eq.~\eqref{eq:defOmega}, we get to the explicit expression
\begin{equation}
\begin{split}
 \omega_k^2 = M (M+2m)  c^4 + k^2 c^2 (1+M/m) + \frac{k^4}{4m^2} ,
\label{disprel}
\end{split}
\end{equation}
in terms of the mass parameter $M$ defined by
\begin{equation}
M c^2 =  E_0 + m c^2  - \omega_p\geq  0.
\end{equation}
The presence of a finite phonon rest energy is a crucial difference as compared to the equilibrium case where phonons are always massless. The phonon mass is, however, dramatically suppressed, $M\ll m$, when the pump frequency approaches resonance with the (interaction-shifted) cavity mode, $\omega_p \simeq E_0 + g |{\mathtt \Phi}_0|^2$, that is, when the operating point approaches the leftmost end point of the upper branch of the bistability loop. In this limit, $M\to 0$ and the dispersion exactly recovers the usual Bogoliubov dispersion of equilibrium Bose condensates~\cite{pitaevskii2003bose}, with massless phonons and a low-frequency speed of sound equal to $c$. 

As usual for quantum Langevin equations, the equation of motion \eqref{eq:2x2eom} also involves a decay term proportional to $\Gamma$ and an effective quantum source term
\begin{equation}
\label{eq:Sdef}
\hat S_\bk(t) \doteq \int d\zeta\, g_\zeta \,\hat c(\bk,\zeta) \, \ep{- i (\omega_\zeta -\omega_p)t}
\end{equation}
summarizing quantum fluctuations in the initial state of the environment, assumed to be decorrelated from the system. In the Markovian limit $\omega\ll \omega_p$, the Langevin quantum noise operator $\hat S_\bk(t)$ satisfies the bosonic commutation relations of a destruction operator 
\begin{subequations}
\begin{align}
 [\hat S_\bk(t) , \hat S^\dagger_{\bk'}(t')] & = \int d\zeta g_\zeta^2 \ep{-i (\omega_\zeta-\omega_p) (t-t')}\\
 \nonumber & = 2 \Gamma \delta(t-t') \delta(\bk - \bk'), \\
 [\hat S_\bk(t) , \hat S_{\bk'}(t')] &= 0.
\label{Scomm}
\end{align}
\end{subequations}
We further assume that the environment is initially in an equilibrium thermal state $\hat \rho_{\rm e}$ with low temperature $T_e \ll \omega_p$. As the characteristic phonon frequencies $\omega_k$ are also much smaller than $\omega_p$, we can safely approximate the expectation values by the following expressions:
\begin{subequations}
\begin{align}
{\rm Tr}\left (\hat \rho_{\rm e}  \hat S_\bk(t)\hat S_{\bk'}(t') \right ) & = 0 , \\
{\rm Tr}\left (\hat \rho_{\rm e} \hat S_\bk (t)\hat S_{\bk'}^\dagger(t')\right ) &\simeq 2\Gamma \, \delta(t-t')\,\delta(\bk-\bk'), \label{eq:whitenoise}\\ 
{\rm Tr}\left (\hat \rho_{\rm e} \hat S_\bk^\dagger (t)\hat S_\bk(t')\right ) &= \frac{2 \Gamma}{\ep{ {\omega_p}/{T_{\rm e}}}-1} \,\delta(t-t')\,\delta(\bk-\bk')\simeq 0 .
\end{align}
\end{subequations}
This means that the environment is a vacuum white-noise bath with a flat frequency distribution. 

\subsection{Quantum fluctuations in the steady state}

In the present stationary case, the Bogoliubov transformation of Eq.~\eqref{eq:bogotf} is time independent. In terms of the phonon operator $\hat \varphi_{\bk}$, Eq.~\eqref{eq:2x2eom} then becomes,
\begin{equation}
\begin{split}
i(\partial_t  + \Gamma  ) \hat \varphi_{\bk} = \omega_k \hat \varphi_\bk + \left (u_k \hat S_\bk  - v_k \hat S_{-\bk}^\dagger\right ).
\label{eq:phoeom}
\end{split}
\end{equation}
Because of $\omega_p > 0$, the creation operator $\hat S_{-\bk}^\dagger$ contains positive frequency. Indeed, using Eq.~\eqref{eq:Sdef}, one gets
\begin{equation}
\begin{split}
\int dt \ep{i \omega t} \hat S_{-\bk}^\dagger &= 2 \pi \int d\zeta\, g_\zeta \,\hat c_0^\dagger(\bk,\zeta) \, \delta (\omega + \omega_\zeta -\omega_p) , 
\label{Spositf}
\end{split}
\end{equation}
which vanishes only for $ \omega > \omega_p$, that is, far outside the frequency range involved in the phonon dynamics. As a result, the quantum fluctuations of the environment heat up the phonon state even when $\hat \Psi^0_\zeta$ is in its vacuum state, i.e., when the environment state is annihilated by $\hat S_\bk$.

The solution of Eq.~\eqref{eq:phoeom} has the following structure:
\begin{equation} 
\begin{split}
\label{Phidecom}
\hat \varphi_\bk(t) &=  \hat \varphi_\bk^{\rm dec}(t;t_0) +  \hat \varphi_\bk^{\rm dr}(t;t_0) .
\end{split} 
\end{equation}
The decaying part is 
\begin{equation}
\label{eq:phidec}
\begin{split}
 \hat \varphi_\bk^{\rm dec}(t;t_0)  = \hat b_\bk \, \ep{- \Gamma (t-t_0)} \ep{- i \omega_k (t-t_0)} ,
\end{split}
\end{equation}
where the $\hat b_\bk$ operator destroys a phonon at time $t_0$ and obeys the canonical commutator $[\hat b_\bk, \hat b_{\bk'}^\dagger] = \delta(\bk - \bk')$. The driven part is 
\begin{equation}
\label{eq:phidr}
 \begin{split}
 \hat \varphi_\bk^{\rm dr}(t;t_0) =  -i \int_{t_0}^t dt'  &\ep{- \Gamma (t-t')} \ep{- i \omega_k (t-t')} \\
 &\times \left [u_k \hat S_\bk (t') - v_k \hat S_{-\bk}^\dagger(t')\right ].
 \end{split}
 \end{equation}
One verifies that $\hat \varphi_\bk$ of \eq{Phidecom} obeys the usual equal time commutators 
\begin{subequations}
\begin{align}
[\hat \varphi_\bk(t), \hat \varphi_{\bk'}^\dagger (t) ] &= \delta(\bk - \bk'), \\
\left[\hat \varphi_\bk(t), \hat \varphi_{\bk'}(t)\right] &=  0,
\end{align}
\end{subequations}
as an identity, irrespective of the choice of $t_0$. More precisely, the two-time commutators are given by
\begin{subequations}
\begin{align}
\label{comm1}
[ \hat \varphi_\bk(t) ,  \hat  \varphi_{\bk'}^\dagger (t') ] &= \ep{- \Gamma\abs{t-t'}} \ep{- i\omega_k (t-t')}\,\delta(\bk - \bk'), \\
[ \hat  \varphi_\bk(t) ,  \hat \varphi_{-\bk'} (t') ] &= \mathcal{O}\left ( \frac{\omega_k}{\omega_p}\right )\,\delta(\bk - \bk'). 
\end{align}
\end{subequations}
In the Markov limit under consideration here, $\omega_k\ll \omega_p$, the latter commutator is negligible. Once the stationary state has been reached (i.e., in the $t_0\to -\infty$ limit), the decaying part $\hat \varphi_\bk^{\rm dec}$ is also negligible and $\hat\varphi_\bk$ is given by $\hat\varphi_\bk^{\rm dr}$ of \eq{eq:phidr}.

\subsubsection{ Two-point functions in the steady state}

The statistical properties of the phonon field are summarized by two point correlation functions 
\begin{equation}
\begin{split}
\label{Gphidef}
G^{\varphi^\dagger \varphi}(t,t'; \bk) & \doteq {\rm Tr}\left ( \hat \rho_0 \hat \varphi_{\bk} ^\dagger (t)\hat \varphi_{\bk} (t') \right ) ,\\
G^{\varphi \varphi}(t,t'; \bk) &\doteq {\rm Tr}\left ( \hat \rho_0 \hat \varphi _{-\bk} (t)\hat \varphi_{\bk} (t')\right) , 
\end{split}
\end{equation} 
which are directly related to the physically observable second-order coherence function, 
\begin{multline}
\label{eq:g1g2def}
g_2 (\xx,t,\xx',t')  \doteq    \\ 
\frac{{\rm Tr} \left (\hat \rho_{0} \, \hat \Phi^\dagger(\xx,t) \hat \Phi^\dagger(\xx',t') \hat  \Phi(\xx',t') \hat \Phi(\xx,t) \right )}{{\rm Tr} \left (\hat \rho_{0} \, \hat \Phi^\dagger(\xx,t) \hat \Phi(\xx,t) \right )\,{\rm Tr} \left (\hat \rho_{0} \, \hat \Phi^\dagger(\xx',t') \hat \Phi(\xx',t') \right ) } ,
\end{multline}
describing the correlations of density fluctuations of the in-cavity photon field. In a typical experiment, this quantity is experimentally accessible by looking at the intensity fluctuations of the emitted radiation from the cavity~\cite{sarchi,Carusotto:2012vz}. The Fourier transform of $g_2 (\xx,t,\xx',t')$ is related to the so-called structure factor of the fluid and provides direct information on the $\bk$ component of the density fluctuations~\cite{pitaevskii2003bose}. To quadratic order in $\delta \hat \Phi$ it is equal to 
\begin{equation}
\begin{split}
g_{2 ,\bk} (t,t')\! \doteq & {\abs{{\mathtt \Phi}_0}^2} \int d(\xx-\xx') \ep{ -i \bk (\xx-\xx')}g_2 (\xx,t,\xx',t')\\
= &  2 \left (u_k-v_k  \right )^2 \Re[G(t,t', \bk)]\\
 &-2 v_k(u_k-v_k) \ep{- \Gamma\abs{t-t'}} \cos[\omega_k (t-t')] ,
\label{g2k}
\end{split}
\end{equation}
where 
\begin{equation}
\begin{split}
G(t,t', \bk) \doteq G^{\varphi^\dagger\varphi}(t,t', \bk) + G^{\varphi \varphi}(t,t', \bk).
\label{RGdef}
\end{split}
\end{equation}
We thus see that a measurement of $g_2$ provides complete information on $\Re[G]$: the term on the last line of \eq{g2k} is in fact state independent, as it is equal to the real part of the commutator in \eq{comm1} multiplied by some known factor. 

Using Eqs.~\eqref{eq:phidr} and~\eqref{eq:whitenoise}, one easily obtains the two-point functions of \eq{Gphidef} in the stationary state:
\begin{equation}
\begin{split} 
\label{Gphieq}
G_{\rm st}^{\varphi^\dagger\varphi}(t,t'; \bk) &= n^b_{k,\rm st}\,\ep{- \Gamma \abs{t-t'}} \ep{ i\omega_k(t-t')},\\
G_{\rm st}^{\varphi\varphi}(t,t'; \bk) &= \bar c^b_{k,\rm st}\, \ep{- (\Gamma +i \omega_k) \abs{t-t'}}  ,\\
\end{split}
\end{equation}
where 
\begin{equation}
\label{eq:nbarceq}
\begin{split}
n^b_{k,\rm st} =v_k^2,\quad \bar c ^b_{k,\rm st} =\frac{ u_k v_k \Gamma}{\Gamma + i \omega_k}.
\end{split}
\end{equation}
Roughly speaking, these quantities give the mean occupation and the correlation function in the phonon point of view. As shown by a more careful analysis, these identifications are subjected to some inherent imprecision; see the discussion below in Sec.~\ref{sec:weakdissip}. 

An alternative description of this state in terms of the photon variables (instead of the phonon ones) can be obtained using the Bogoliubov transformation Eq.~\eqref{eq:bogotf}. The mean occupation number and the correlations of photon operators are equal to
\begin{equation}
\begin{split}
n_k^{a,{\rm st}} &= 2 u_k^2 v_k^2 \frac{\omega_k^2}{\Gamma^2+\omega_k^2} , 
\label{eq:nka}\\
c_k^{a,{\rm st}} &= i\omega_k u_k v_k \left[ \frac{v_k^2 }{ \Gamma -i \omega_k}- \frac{u_k^2}{ \Gamma +i \omega_k}   \right].
\end{split}
\end{equation}
These quantities are accessible from the intensity pattern of the far-field emission from the cavity and its coherence properties: the presence of a non-vanishing emission $n_k^{a,{\rm st}}$ at a wavevector distinct from the coherent pump at $\bk=0$ stems from parametric processes analogous to the ones taking place in parametric down-conversion experiments. The non-vanishing correlation $c_k^{a,{\rm st}}\neq 0$ is a signature of the two-mode squeezed nature of this emission~\cite{d1995quantum,gardiner2004quantum,PhysRevB.63.041303}.

From the photon momentum distribution Eq.~\eqref{eq:nka}, it is immediate to calculate the first-order coherence function defined as 
\begin{equation}
\label{eq:g1def} 
\begin{split}
g_1 (\xx,t,\xx',t')  \doteq {\rm Tr} \left (\hat \rho_{0} \, \hat \Phi^\dagger(\xx,t)\,\hat \Phi(\xx',t') \right ).
\end{split}
\end{equation}
For simplicity, we restrict our attention to $g_1$ evaluated at equal times $t=t'$,
\begin{equation}
\begin{split}
g_1 (\xx,\xx',t'=t) =\abs{{\mathtt \Phi}_0}^2+\int \! \frac{d\bk}{(2\pi)^d}\, e^{i\bk\cdot(\xx'-\xx)}\,n_\bk^{a,{\rm st}}.
\label{g1anal}
\end{split}
\end{equation}
The modified Bogoliubov coefficients $u_k, v_k$ which appear in \eq{eq:nka} are given in \eq{eq:bogotf} and the frequency $\omega_k$ in Eq.~\eqref{disprel}. Using these expressions, a straightforward calculation gives
\begin{equation}
n_k^{a,{\rm st}} = \frac{m^2 c^4}{2(\Gamma^2+\omega_k^2)}
\end{equation}
which is regular in the $k\to 0$ limit both because of the (small) phonon mass $M(M+2m)c^4$ in \eq{disprel} and because of losses. 

Using isotropy, it is immediate to see from the preceding discussion that for any dimensionality $d$, the $|\xx-\xx'|\to\infty$ long distance limit of $g_1$ shows a condensate plus an exponentially decaying term,
\begin{equation}
g_1 (\xx,\xx',t'=t) \simeq \abs{{\mathtt \Phi}_0}^2 + A \, e^{-|\xx-\xx'|/\ell_c} ,
\end{equation}
with a coherence length
\begin{equation}
\ell_c=\left[\frac{c^2(1+M/m)}{\Gamma^2+M(M+2m)c^4}\right]^{1/2}.
\end{equation}
This shows that in the present case, thanks to the presence of the coherent pump, the long-distance coherence of the photon ``condensate" is robust against fluctuations independently of the dimensionality. For a recent discussion of the long distance coherence of $g_1$ under an incoherent pump, we refer to~\cite{Chiocchetta:arXiv1302.6158,Altman:arXiv1311.0876}. 

\subsubsection{Dissipationless limit}

\begin{figure*}
\begin{minipage}{0.45\linewidth}
\includegraphics[width=1\linewidth]{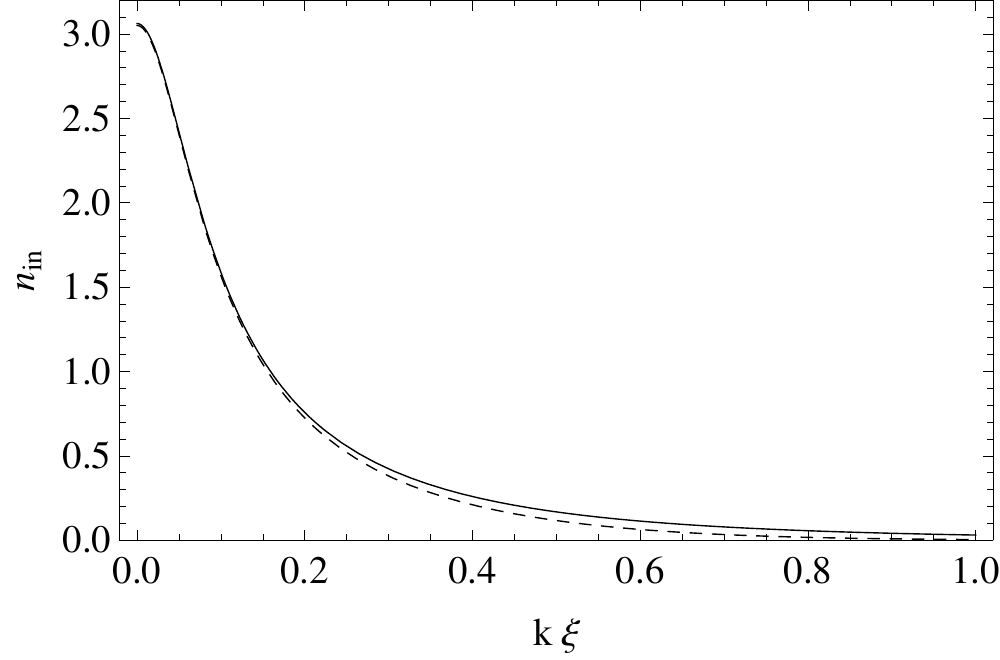}
\caption{Mean occupation number of phonons $n_{k,\rm st}^b = v_k^2$  for a low phonon mass parameter $M/m = 0.01$ in the driven-dissipative steady state (solid), and in a thermal state at temperature $T= 1/2 m c^2 $ (dashed). Both curves are independent of the dissipative rate $\Gamma$. Whatever the value of the mass parameter $M$, one can show that the {\em absolute} deviation between the two curves is always smaller than $0.052$ which is reached for $\Omega_k \sim 1.5 m c^2$. On the other hand, the {\em relative} difference between the two occupation numbers becomes large at high momenta; see Fig.~\ref{fig:nandbarc}.} 
\label{fig:ninofk}
\end{minipage}
\hspace{0.04\linewidth}
\begin{minipage}{0.45\linewidth}
\includegraphics[width=1\linewidth]{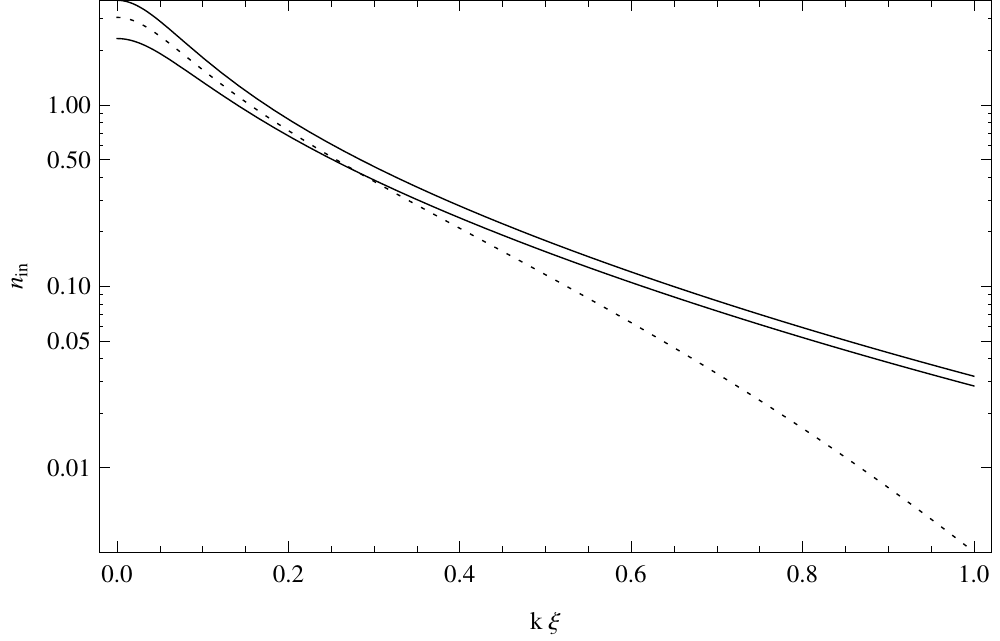}
\caption{ Intrinsic imprecision in the measurement of the mean occupation number of phonons. The solid curves show $ n_k^b= n_{k,\rm st}^b \pm \abs{\bar c_{k,\rm st}^b}$ involved in \eq{RG4}. The dotted line represents the thermal distribution at $T= 1/2 m c^2 $: while $n_k^b$ decays as $1/k^2$ for large momenta in the non-equilibrium stationary state, the thermal distribution decays according to a much faster Boltzmann law as $\ep{- (k /mc)^2 }$. The system parameters are the same as in Fig.~\ref{fig:ninofk}: phonon mass $M/m = 0.01 $ and dissipation rate $\Gamma/m c^2 = 0.03 $.}
\label{fig:nandbarc}
\end{minipage}
\end{figure*}

To understand the physical implications of these results, we first consider the $\Gamma \to 0$ limit where dissipation tends to zero. Note that because of the presence of the pump the system does not recover a standard thermodynamical equilibrium state in this limit, but maintains a non equilibrium character. More details on this crucial fact are given in Appendix~\ref{app:FD}. In this case, the effective phonon state is incoherent, since $G_{\rm st}^{\varphi\varphi} =  \bar c ^b_{k,\rm st} = 0$, as in standard thermal equilibrium. The state is thus fully characterized by the finite value of Eq.~\eqref{eq:nbarceq} of the mean phonon occupation number $n^b_{k}$. Interestingly, even in the $\Gamma \to 0$ limit, the stationary state of the system differs from a standard thermodynamical equilibrium state, as is manifest in the phonon occupation distribution not following the Planck distribution. Nonetheless, as can be seen in Fig.~\ref{fig:ninofk}, the state it is very close to a thermal state at temperature $k_B T_{\rm st} = m c^2/2 $ fixed by the interaction energy; see \eq{inten}. This is our first result: Because of the unusual presence of positive frequency in  $\hat S_{-\bk}^\dagger$ [see \eq{Spositf}], the phonon field is effectively heated up even when the environment is in its vacuum state. More details on the non equilibrium origin of this crucial fact are given in Appendix~\ref{app:FD}. In Fig.~\ref{fig:ninofk} and in subsequent figures, the wave vector $k$ is adimensionalized by making use of the healing length defined by $\xi = 1/ 2m c $ (since $\hbar = 1$). 

From a physical point of view, it is important to note the conceptual difference of this result with respect to the quantum depletion of the Bose condensate as predicted for the ground state of equilibrium Bogoliubov theory~\cite{pitaevskii2003bose}: the finite occupation number $n^b_{k}$ refers here to phonon quasi-particle excitations, while standard quantum depletion refers to the underlying particles (in our case, photons). Along the same lines, one should not confuse the finite phonon occupation in the present driven-dissipative stationary state, with the finite photon occupation in the ground state of a microcavity device in ultra strong light-matter coupling, as discussed in~\cite{bastard,PhysRevA.74.033811}.

To better understand the physical meaning of the two different photon and phonon descriptions of the same state, it is useful to introduce the concept of non-separability of the quantum state. For generic homogeneous states, unitarity implies the inequality
\begin{equation}
\begin{split}
\abs{c_k}^2 \leq n_k(n_k+1),
\end{split}
\label{eq:unita}
\end{equation}
while separability of the state~\cite{Werner:1989} imposes the stronger condition  
\begin{equation}
\label{eq:separability}
\begin{split}
\abs{c_k} \leq n_k ;
\end{split}
\end{equation}
see Appendix~\ref{app:CS} for more details. In our system, \eq{eq:separability} can be applied in two distinct ways, either to photon or to phonon operators: the results are not expected to coincide as photon and phonon operators are related by \eq{eq:bogotf} which is a $U(1,1)$ transformation mixing creation and destruction operators. 

From the phonon point of view, the stationary state of the system is manifestly separable in the $\Gamma\to 0$ limit as phonons are fully incoherent, $\bar c ^b_{k,\rm st} = 0$. On the contrary, the same state is non-separable from the photon point of view since 
\begin{equation}
\abs{c_k^a}^2 = n_k^a( n_k^a + 1/2) ,
\label{photCS}
\end{equation}
violates the separability bound \eq{eq:separability}.

Even though the state is non separable only from the photon point of view, we can explicitly verify that the entropy of the state agrees in the two points of view, as is expected from the invariance of entropy under $U(1,1)$ transformations. This is straightforwardly done knowing that the entropy is equal to 
\begin{equation}
 S= 2[(\bar n +1) \log (\bar n +1) - \bar n \log \bar n ],
\label{Sent}
\end{equation}
in terms of $\bar n$ defined by $(\bar n +1/2)^2 \doteq (n+1/2)^2 - \abs{c^2}$~\cite{Campo:2005sy}.

\subsubsection{Weak dissipation} 
\label{sec:weakdissip}

Having understood the state properties in the limit $\Gamma \to 0$, we now turn to the case of small but finite dissipative rates, $\Gamma \ll\omega_k$. The main change is that the correlation function \eq{Gphieq} is now $G_{\rm st}^{\varphi \varphi} \neq 0$. While its $t-t'$ time dependence correctly expresses stationarity of the state, it dramatically differs from the usual $\omega_k(t+t')$ one describing correlations of real phonon pairs at wave vectors $\pm \mathbf{k}$~\cite{Carusotto:2009re,Busch:2013sma}. This means that $G_{\rm st}^{\varphi \varphi} \neq 0$ {\it cannot} be straightforwardly interpreted as describing {\em real} pairs of phonons with opposite momenta. Still, because of the quantum fluctuations associated with the dissipation processes, there are non-trivial correlations $G_{\rm st}^{\varphi \varphi} \neq 0$ between phonon modes of opposite wave vectors $\pm\bk$. This is a second main result of this paper.

The presence of a non-zero correlation $\bar c ^b_{k,\rm st}\neq 0$ in the stationary state has important consequences when one attempts to measure the occupation number $n^b_{k,\rm st}$ via a measurement of $g_2$ and thus of $\Re[G]$. To be specific, let us consider an experiment where $\Re[G(t,t', \bk)]$ is measured for various values of the interval $\tau=t'-t$. Provided $\tau$ is short enough, $\Gamma \tau \ll 1$, one gets
\begin{equation}
\begin{split}
\Re[G_{\rm st}(t,t+\tau, \bk)] \sim n^b_{k,\rm st} \cos(\omega_k \tau) + \Re (\bar c^b_{k,\rm st}\ep{- i \omega_k \tau }).
\label{RG3}
\end{split}
\end{equation}
For very small dissipation rates $\Gamma \to 0$, correlations are negligible: as a result, the left-hand side divided by $\cos( \omega_k \tau)$ directly provides information on the mean number of particles $n^b_{k, \rm st}$. When we proceed in the same way in the presence of a significant dissipation, the same procedure gives 
\begin{equation}
\begin{split}
\tilde{n}^b_{k, \rm st} = n^b_{k,\rm st}  + \Re (\bar c ^b_{k,\rm st}) + \Im (\bar c^b_{k,\rm st}) \tan( \omega_k \tau), 
\label{RG4}
\end{split}
\end{equation}
which shows periodic deviations in $\tau$ around an average value $ n^b_{k,\rm st}  + \Re (\bar c ^b_{k,\rm st})$; note that this average still differs from $n^b_{k,\rm st}$ by a systematic error proportional to $\bar c ^b_{k,\rm st}$ (see Fig.~\ref{fig:nandbarc}).

\section{Phonon pair production by a sudden modulation} 
\label{sec:DCE}

In the previous section, we studied the quantum fluctuations in a stationary state under a monochromatic continuous wave pump. In this section we shall extend the discussion to the case when a sudden change is imposed on the system and pairs of phonons are expected to be generated at the time of the fast modulation via processes that are closely analogous to the cosmological pair creation effect in the early universe~\cite{Fedichev:2003bv,Campo:2003pa,Adamek:2013vw} and to the dynamical Casimir effect~\cite{Carusotto:2009re,PhysRevLett.109.220401}. 

\subsection{The modified state} 

To facilitate analytical calculations, we will restrict our attention here to a very idealized model inspired by Ref.~\cite{Carusotto:2008ep}, where the spatially homogeneous condensate wavefunction of amplitude $\mathtt{\Phi}_{0} $ remains an exact solution of Eq.~\eqref{eq:GPEomegae} at all times. As compared to atomic gases, this requirement is slightly more subtle in the present non equilibrium case as the photon density is related to the pump intensity by the more complicated state equation \eqref{eq:GPEhomo}. A possible strategy to fulfill this condition might consist of assuming that $\Gamma$, $m$, the pump amplitude $F_0$, and its frequency $\omega_p$ remain constant while $g$ and $E_0$ suddenly change at $t = 0$, keeping $E_0(t) +  g(t)\abs{ \mathtt{\Phi}_{0}}^2$ constant.\footnote{
In this case, the change is specified by one parameter. Two-parameter changes can be considered by changing both $\Gamma$ and $F_0$ while keeping their ratio constant. This can still be generalized by changing both $\omega_p$ and $E_0$ while keeping $\omega_p - E_0$ constant. In all cases, the gluing of the background across the jump is easily done.} 
While we agree that such modulations are quite unrealistic in state-of-the art experiments, still the predicted phonon pair production process appears to be conceptually identical to the one taking place in the more realistic but more complex configurations where the condensate wave function is itself varying, as in, e.g., Ref.~\cite{Koghee2013}.

As a result of the modulation, the phonon frequency $\omega_k(t)$ of Eq.~\eqref{eq:omk} experiences a sudden change (the subscript $\pm$ refers to its value at times $t \gtrless 0$)
\begin{equation}
\label{eq:omegajump}
\begin{split}
\omega_k(t) &= \omega_{k, -} + \theta(t) (\omega_{k, +} - \omega_{k, -}) ,
\end{split}
\end{equation}
which directly reflects onto the Bogoliubov operators: while the photon operator $\hat \phi$ in Eq.~\eqref{eq:2x2eom} is continuous at $t=0$, the phononic ones $\hat \varphi$ defined in Eq.~\eqref{eq:bogotf} experience the following sudden jump~\cite{Carusotto:2009re,Busch:2013sma}
\begin{equation}
\label{eq:bogojump}
\begin{split}
\hat \varphi_{\bk,+} &= \alpha_{k} \hat \varphi_{\bk,-} + \beta_{k}\hat \varphi^\dagger_{-\bk,-} ,\\
\alpha_{k}& = u_{k,+} u_{k,-}-v_{k,+} v_{k,-} = \frac{\omega_+ \, +\,  \omega_-}{2\sqrt{\omega_+ \omega_-}}, \\ 
\beta_{k} &= v_{k,+} u_{k,-} -u_{k,+} v_{k,-}= \frac{\omega_+ \, -\,  \omega_-}{2\sqrt{\omega_+ \omega_-}},
\end{split}
\end{equation} 
where the second equalities follow from the constancy of $\Omega_k-mc^2$. 

Hence, for positive times $t$ and with, we have $ \hat \varphi_\bk^{\rm dec}(t) $ given by Eq.~\eqref{eq:phidec} with $t_0 =0^+$, $\hat b_\bk= \hat \varphi_{\bk,+} $, and $ \hat \varphi_\bk^{\rm dr}(t) $ given by Eq.~\eqref{eq:phidr}. Using the fact that the source term $\hat S$ has a white noise profile, at all times $t>0$ one has
\begin{equation}
\begin{split}
\label{G0}
\mathrm{Tr}\left ( \hat \rho_0 \hat \varphi_\bk^{\rm dec} \hat \varphi_{-\bk}^{\rm dr}   \right ) = \mathrm{Tr}\left ( \hat \rho_0 \hat \varphi_\bk^{\rm dec} (\hat \varphi_{\bk}^{\rm dr})^\dagger   \right )=0.
\end{split}
\end{equation}
For $t,t'>0$, after the jump, the two-point correlation functions defined in \eq{Gphidef} then have the forms
\begin{equation}
\label{eq:diisnb}
\begin{split}
G^{\varphi^\dagger,\, \varphi}_{\rm DCE}(t,t', \bk)& \! =  \! \left ( n_{k,\rm f}^b \ep{- \Gamma \abs{t-t'}} + \delta n_{k}^b  \ep{- \Gamma (t+t')} \right ) \! \ep{ i \omega_{k,+}(t-t')}\\
G^{\varphi,\, \varphi}_{\rm DCE}(t,t', \bk)& \! = \!  \bar c_{k,\rm f}^b \ep{- (\Gamma +i \omega_{k,+}) \abs{t-t'}} + \! c_k^b  \ep{- (\Gamma + i \omega_{k,+})(t+t')} .
\end{split}
\end{equation}

\begin{figure}
\includegraphics[width=1\linewidth]{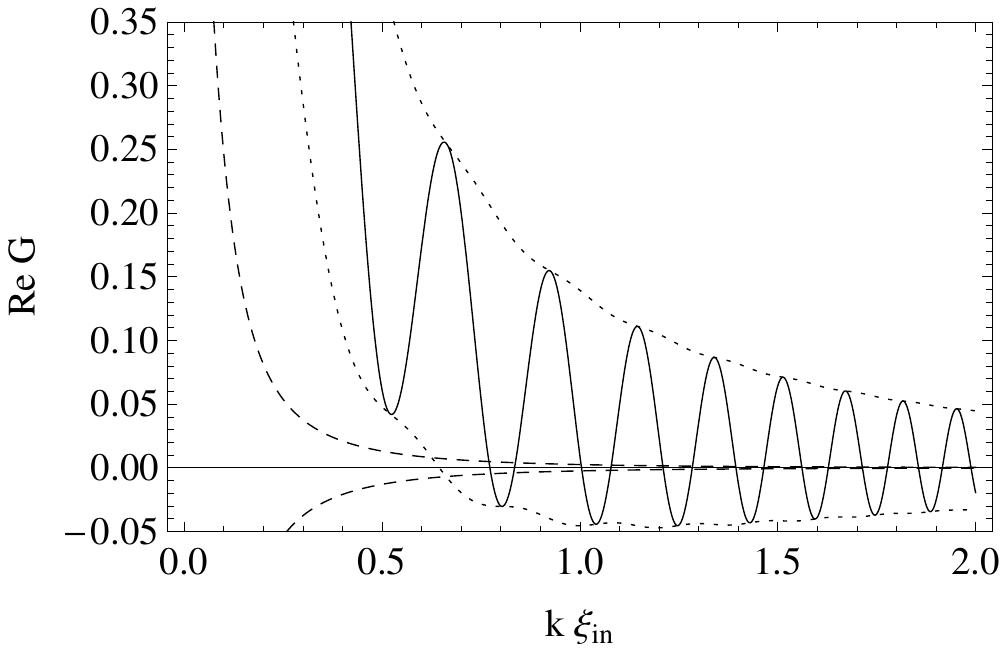}
\caption{Separability of the phonon state after a sudden jump. The oscillating solid line shows the equal-time $\textrm{Re}[G_{\rm DCE}(t,t,\mathbf{k})]$ defined in \eq{RG2} for $t =3/m c_{\rm in}^2$ as a function of (normalized) phonon momentum $\xi_{\rm in} k$. Non separable phonon states are found wherever the lower envelope (dotted line), indicating $G_{DCE}^{\varphi^\dagger\varphi}(t,t,\bk)-|G_{DCE}^{\varphi\varphi}(t,t,\bk)|$, goes below $0$. In the present case, the intrinsic imprecision $\pm |{\bar c_{k,\rm f}^b}|$ (dashed lines) does not significantly affect the identification of non separable states.  System and jump parameters: $M_{\rm in}/{ m } = 0.01$, $c_{\rm f}^2/c_{\rm in}^2  = 2$, and $\Gamma = 0.03 m c_{\rm in}^2$. }
\label{fig:nkmck}
\end{figure}

Four independent and constant quantities are identified in \eq{eq:diisnb} through the time dependence of their associated exponential factor, namely,
\begin{equation}
\begin{split}
\label{4q}
n_{k,\rm f}^b   &=v_{k,+}^2,  \\ 
\bar c_{k,\rm f}^b &= \frac{u_{k,+} v_{k,+} \Gamma}{\Gamma + i \omega_{k,+}}, \\
\delta n_{k}^b&= {\rm Tr}\left [ \hat \rho_0 \, \hat \varphi_{\bk,+}^\dagger\, \hat \varphi_{\bk,+} \right ] - n_{k,\rm f}^b , \\
c_k^b &={\rm Tr}\left [ \hat \rho_0\,  \hat \varphi_{-\bk,+}\, \hat \varphi_{\bk,+} \right ]- \bar c_{k,\rm f}^b .
\end{split}
\end{equation}
The first two quantities $n_{k,\rm f}^b$ and $\bar c_{k,\rm f}^b$ give the final values once the stationary state is again reached for the new parameters after the jump: they have the same physical interpretation as $n_{k,\rm st}^b$ and $\bar c_{k,\rm st}^b$ defined in Eq.~\eqref{eq:nbarceq} and discussed at length in the previous section. Instead, $\delta n_{k}^b$ and $c_k^b$ govern the time dependence of the correlation functions in response to the jump in the parameters. They involve two traces taken at a time $t=0^+$ which are [see Eq.~\eqref{eq:bogojump}] 
\begin{equation}
\begin{split}
 {\rm Tr}\left[ \hat \rho_{0}  \hat \varphi_{\bk,+}^\dagger\, \hat \varphi_{\bk,+} \right]  &=  \left (\alpha_{k} ^2 +\beta_{k} ^2\right )  n_{k,\rm in}^b+\beta_{k} ^2 \\
&\quad + 2\alpha_{k}  \beta _{k} \Re\left ( \bar c_{k, \rm in}^b \right ) , \\
{\rm Tr}\left [ \hat \rho_{0}  \hat \varphi_{-\bk,+}\, \hat \varphi_{\bk,+}\, \right ]&=\alpha_{k} ^2 \bar c_{k, \rm in}^b+\beta_{k} ^2 (\bar c_{k,\rm in}^b)^*\\
&\quad +\alpha_{k}  \beta_{k}  \left(2 n_{k,\rm in}^b+1\right), \\
\end{split}
\end{equation}
where
\begin{equation}
\begin{split}
n_{k,\rm in}^b= v_{k,-}^2 ,\quad \bar c_{k,\rm in}^b = \frac{\Gamma u_{k,-} v_{k,-}}{\Gamma + i \omega_{k,-}}
\end{split}
\end{equation}
are the initial stationary values as predicted by Eq.~\eqref{eq:nbarceq}. More specifically, $\delta n_k^b$ is involved in the only decaying term in Eq.~\eqref{eq:diisnb} which oscillates, $\ep{i\omega_{k,+} (t-t')}$: physically, its equal-time value $\delta n_k^b(t) = \delta n_k^b \, \ep{- 2 \Gamma t}$ describes the number of extra photons with respect to $n_{k,\rm f}^b$ that are generated by the jump and still present at time $t$. $c_k^b$ is instead involved in the only term which is rotating as $\ep{-i \omega_{k,+} (t+t')}$: its equal-time value $c_k^b(t) = c_k^b \, \ep{- 2 (\Gamma+i\omega_{k,+}) t}$ gives the instantaneous correlation between these extra phonons: as it is illustrated in Fig.~\ref{fig:nkmck}, these non-trivial correlations can produce nonseparability at the level of phonons. Studies of this physics for lossless systems were reported in~\cite{Campo:2005sy,Campo:2005sv,Campo:2008ju,Bruschi:2013tza,Busch:2013sma,finazzi2013}.

\begin{figure*}
\begin{minipage}{0.45\linewidth}
\includegraphics[width=1\linewidth]{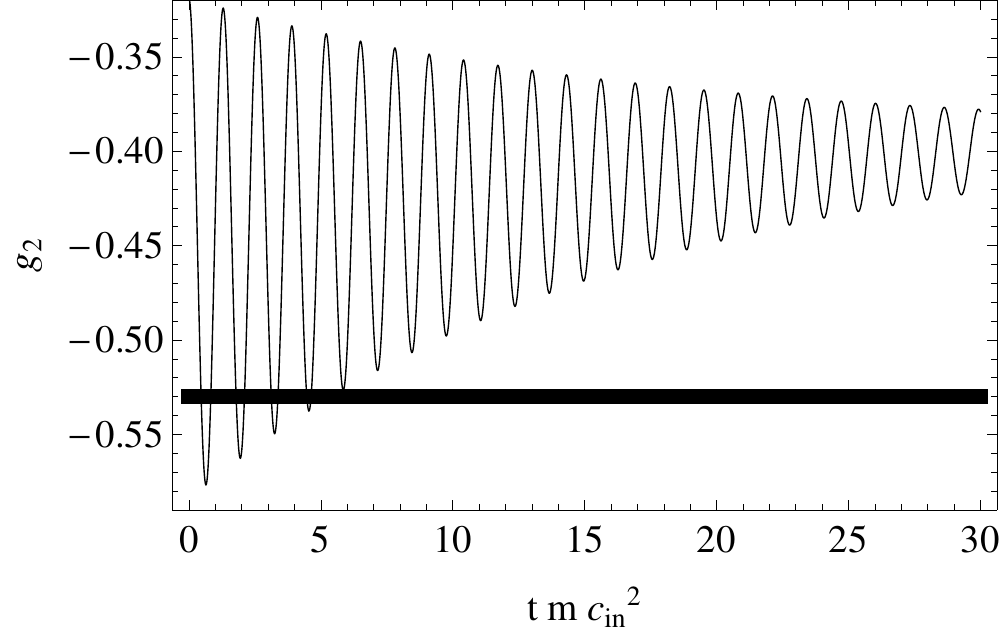}
\caption{Time evolution of the Fourier-space second-order coherence of the photon field. The solid line shows the equal-time $g_{2, \, \bk}(t,t')$ at $t=t'$ for a given $k \xi_{\rm in} = 0.75 $. In this case, the initial and final values of the phonon occupation are respectively $n^b_{{\rm st},k}\simeq 0.06$ and $n^b_{{\rm f},k}\simeq 0.14$. An exponential convergence towards the final value is apparent. The horizontal line represents the phonon separability threshold: Non separability is found as long as the lower envelope of the oscillating solid line stays below the horizontal line, whose thickness shows the intrinsic imprecision $\pm\abs{\bar c_{k,\rm f}^b}$ of the mean occupation number. The system and jump parameters are the same as in Fig.~\ref{fig:nkmck}.}
\label{fig:g2oft}
\end{minipage}
\hspace{0.04\linewidth}
\begin{minipage}{0.45\linewidth}
\includegraphics[width=1\linewidth]{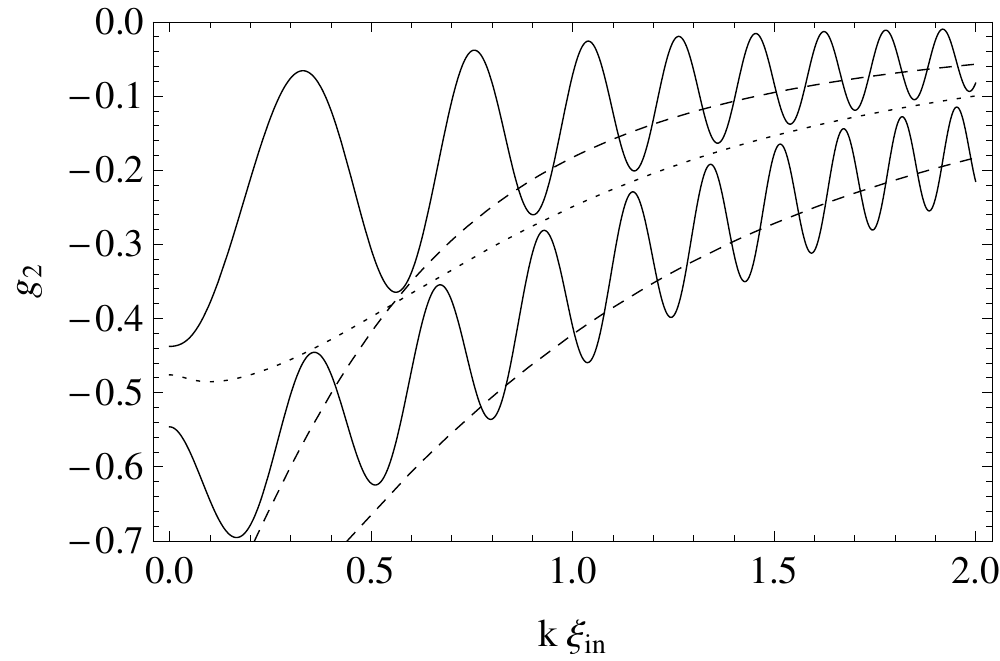}
\caption{Fourier-space second-order coherence of the photon field. The two oscillating curves represent the equal-time $g_{2, \, \bk}(t,t)$ as a function of normalized momentum $k \xi_{\rm in}$ at a time $t = 3/ m c_{\rm in}^2$ after a jump characterized by $c_{\rm f}^2/ c_{\rm in}^2 = 2$ for the upper curve, and $c_{\rm f}^2/ c_{\rm in}^2 = 1/2$ for the lower one. The two dashed curves give the separability thresholds of these two cases: nonseparability is found whenever the lower envelope (not represented here) of an oscillating solid line goes below the corresponding dashed curve. The middle dotted curve is the (common) value of $g_{2,\, \bk}$ before the sudden change. The system and jump parameters are the same as in Fig.~\ref{fig:nkmck}.}
\label{fig:g2ofk}
\end{minipage}
\end{figure*}

For completeness, it is useful to give explicit expression of the corresponding quantities in the photon (rather than phonon) point of view. Using again the Bogoliubov transformation Eq.~\eqref{eq:bogotf}, one gets
\begin{subequations}
\begin{align}
\begin{split}
n_k^a(t) = n_{k,+}^{a,\rm st} + &e^{-2\Gamma t} \left\{ (u_{k,+}^2 + v_{k,+}^2)\,\delta n_k^b+  \right .\\
&  \left. - 2 u_{k,+}v_{k,+} \textrm{Re}[c_b e^{-2i\omega_{k,+} t}]  \right\} ,
\end{split}\\
\begin{split}
c_k^a(t)= c_{k,+}^{a,\rm st} + &e^{-2\Gamma t} \left\{ u_{k,+}^2 c_b e^{-2i\omega_{k,+} t} + \right. \\  
& \left. + v_{k,+}^2 c_b^* e^{2i\omega_{k,+} t}-2 u_{k,+}v_{k,+} \delta n_k^b \right\}.
\end{split}
\end{align}
\end{subequations}
As expected, the Bogoliubov transformation is responsible for temporal oscillations in these photonic quantities in response to the jump: as compared to the atomic case~\cite{Carusotto:2009re}, oscillations are now damped at the loss rate $\Gamma$ and tend to their static values $n_{k,+}^{a,\rm st}$ and $c_{k,+}^{a,\rm st}$ for the final parameters after the jump.

\subsection{The observables}

This last section is devoted to a discussion of possible strategies aiming to extract the four quantities (i.e., six real quantities) of \eq{4q} from accurate measurements of the coherence functions of the cavity photon field. Two of them, $n_{k, \rm f}^b, c_{k, \rm f}^b,$ characterize the final stationary state, while $ \delta n_k^b, {c_k^b}$ characterize the decaying properties of the state. Knowledge of the four real quantities $|{c_k^b}|, \delta n_k^b, n_{k, \rm f}^b$ and $|{\bar c_{k,\rm f}^b}|$ allows assessment of the nonseparability of the phonon state.

As a first example, we consider the equal-time combination analogous to Eq.~\eqref{RGdef},
\begin{multline}
 \Re[G_{\rm DCE}(t,t, \bk)]  =  n^b_{k,\rm f} + \delta n_k^b\,e^{-2\Gamma t} + \\ + \Re[c_k^b\,e^{2 (\Gamma+i\omega_{k,+}) t}+ \bar c_{k,\rm f}^b]. 
\label{RG2}
 \end{multline} 
For underdamped phonon modes such that $\omega_{k,+}>\Gamma$, this quantity oscillates between maxima and minima given by $n^b_{k,\rm f} + \delta n^b_{k}(t) +  \Re[\bar c_{k,\rm f}^b ] \pm \abs{c_k^b(t)}$: the function $c_k^b(t) = c_k^b e^{- 2\Gamma t}$ can thus be extracted from the amplitude of oscillations. The mid-point of the oscillations provides instead information on $n_k^b(t) = n^b_{k,\rm f}+ \Re(\bar c_{k,\rm f}^b)+ \delta n_{k}^b e^{- 2\Gamma t} $. This quantity can be taken as an operative definition of the mean occupation number. If one wishes to extract the $\delta n_{k}^b(t) = \delta n_{k}^b e^{- 2\Gamma t} $ contribution from the correlation correction, one has just to measure $\Re[G_{\rm DCE}(t,t)]$ for different times $t$: since it is the only term that possesses this time dependence, $\delta n_{k}^b(t)$ is therefore well defined. Hence, the only quantity affected by $\bar c_{k,\rm f}^b$ is $n_{k,\rm f}^b$.

In terms of $\Re[G_{\rm DCE}(t,t, \bk)]$, the non-separability condition of Eq.~\eqref{eq:separability} applied to phonon states, $\abs{c_k^b(t)}\geq n_k^b(t)$, is simply reexpressed as $\Re[G_{\rm DCE}(t,t, \bk)] \leq 0$.\footnote{
Notice that $\Re[G_{\rm DCE}(t,t', \bk)]$ coincides with the quantity $\omega_{\rm f} G_{ac}(t, t' , \bk) - 1/2 $ involving the anti-commutator which is used in \cite{Busch:2013sma} to assert the nonseparability of the phonon state.}
up to error terms of order $\mathcal{O}[\bar c_{k, {\rm f}}^{b}]$. The simplicity of this condition arises from the fact that $G_{\rm DCE}(t,t, \bk)$ is the expectation value of normal ordered products of phonon operators $\hat b_\bk, \hat b_{-\bk}^\dagger$ of \eq{eq:phidec}. The condition and its intrinsic imprecision are visually represented in Fig.~\ref{fig:nkmck} by the two dashed lines.

In practice, optical measurements typically involve the coherence function of a field. In our case, the second-order coherence $g_2$ is most important as it is the simplest to analyze. Inserting the expectation values of Eq.~\eqref{eq:diisnb} into Eqs.~\eqref{eq:g1g2def} and~\eqref{g2k}, we immediately identify the stationary and the decaying contributions,
\begin{equation}
\begin{split}
g_{2 ,\bk} (t,t') &= \ep{- \Gamma\abs{t-t'}} g_{2 ,\bk}^{\rm st} (t,t') + \ep{- \Gamma(t+t')} g_{2 ,\bk}^{\rm dec} (t,t').
\end{split}
\label{g2k_split}
\end{equation}
The time dependences of $g_{2 ,\bk}^{\rm st}$ and $g_{2 ,\bk}^{\rm dec}$ is of the form 
\begin{subequations}
\label{eq:g2ofA}
\begin{align}
g_{2 ,\bk}^{\rm st} (t,t')&=A_1 \cos\left [ \omega_{k,+}\abs{t-t'} + \theta_1 \right ],\\
\begin{split}
g_{2 ,\bk}^{\rm dec} (t,t') &= A_2  \cos\left [ \omega_{k,+}(t-t') \right ]\\
 &+A_3 \cos\left [ \omega_{k,+}(t+t') + \theta_3 \right ] ,
 \end{split}
\end{align}
\end{subequations}
where the three constants are
\begin{equation}
\begin{split}
A_1 \ep{- i \theta_1} &= 2(u_k-v_k)^2 \left [n_{k,\rm f}^b + \bar c_{k,\rm f}^b \right ] - 2 v_k (u_k-v_k) , \\
A_2&= 2(u_k-v_k)^2 \delta n_{k}^b  ,\\ 
A_3 \ep{- i \theta_3} &=2(u_k-v_k)^2 c_k^b  . \\ 
\end{split}
\end{equation}
From measurements of $g_{2 ,\bk} (t,t')$ at different times $t,t'$, we can thus extract five real quantities (out of the six physical ones), namely, $\Re[c_k^b], \Im[c_k^b], \delta n_{k}^b, \Im \bar c_{k,\rm f}^b$, and $n_{k,\rm f}^b +\Re[ \bar c_{k,\rm f}^b]$. To disentangle $n_{k,\rm f}^b $ from $\Re[\bar c_{k,\rm f}^b]$, another observable, such as the $\bk$ component of the $g_{1}$ function, is needed. 

In Fig.~\ref{fig:g2oft} we represent the equal-time $g_{2 ,\bk}$ as a function of $t$, for a given wave number $k \xi_{\rm in} =0.75$ and the same parameters as in the previous figures: for these values, the initial value oscillates with amplitude $A_3 = 0.26$ around the mean value $A_1 \cos(\theta_1) + A_2 = -0.45 $. Its final value is $A_1 \cos(\theta_1) = -0.4$. The threshold value of nonseparability is reached when the minimum of the $g_2$ crosses $ -0.53 \pm 0.005$. Neglecting for simplicity the intrinsic imprecision due to $\pm |\bar{c}^b_{k,\rm f}|$, as in~\cite{Busch:2013sma}, losses make nonseparability disappear within a time of the order of
\begin{equation}
\begin{split}
t_{\rm loss} \doteq \log\left ( (\abs{c_k^b}-\delta n_k)/n_{k,\rm f}  \right )/2\Gamma \lesssim 1/4 n_{k,\rm f} \Gamma ,
\label{tloss}
\end{split}
\end{equation}
where the last inequality follows from Eq.~\eqref{eq:unita} and applies when $2n_{k,\rm f}\gg 1$. In the present case, $t_{\rm loss} \Gamma \simeq 0.16 $. 

In Fig.~\ref{fig:g2ofk}, we represent the $k$ dependence of the equal time $g_{2, \, \bk}$ function, at a time $t= 3 / m c_{\rm in}^2$ in two different cases: when ${c_{\rm f}^2}/{c_{\rm in}^2}  = 2$ as in the former figure, but also when ${c_{\rm f}^2}/{c_{\rm in}^2}  = 1/2$, i.e., when the final sound speed is divided by 2 rather than multiplied by 2. In both cases, we use the same system parameters as in the former figure. We observe two oscillating functions, the minima of the upper one close to the maxima of the lower one. Their common value is $g^{\rm st}_{2, \, \bk}$ evaluated before the jump; see Fig.~6 in Ref.~\cite{Busch:2013sma} for more details. It is represented by a dotted line. From the two envelopes of each curve, we can measure the width on the oscillations $A_3 \ep{- 2 \Gamma t}$, which gives the $k$ dependence of the strength of the correlations, and the average value $A_1 \cos(\theta_1) + A_2$. The two dashed curves in Fig.~\ref{fig:g2ofk} are the corresponding thresholds of nonseparability. In both cases, there is a large domain of $k$ where the state is nonseparable. 

\begin{figure}
\includegraphics[width=1\linewidth]{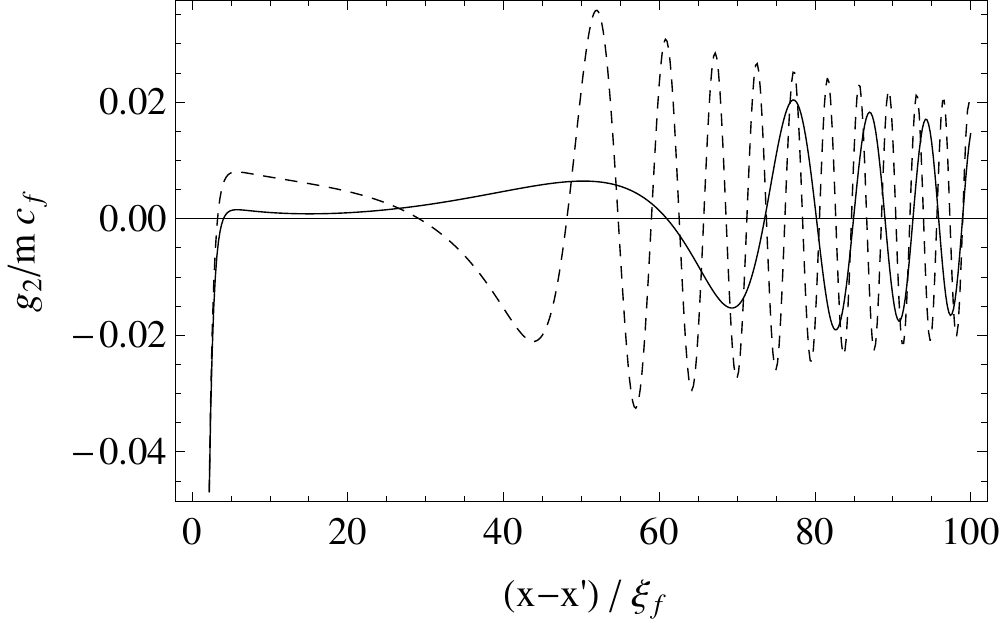}
\caption{Real-space second-order coherence of the photon field in the one-dimensional case. We plot the equal-time $g_2(t,t, x-x')$ as a function of the (normalized) spatial distance $x-x'$ at $t= 12/ mc_{\rm in}^2$ (dashed), and $18/mc_{\rm in}^2$ (solid). In addition to the negative peak at $x=x'$ due to repulsive interactions, the phonon pairs generated at the jump are visible in the series of moving fringes with spatially decreasing spatial period. Given the small value of $\Gamma/m c^2_{\rm in} = 0.03$, dissipative effects have a minor effect on the profiles shown here. Separability features are hard to ascertain from this figure. The system and jump parameters are the same as in Fig.~\ref{fig:nkmck}.}
\label{fig:g2ofx}
\end{figure}

To complete the study of the $g_2$, we represent in Fig.~\ref{fig:g2ofx} its spatial dependence on $\xx-\xx'$ after integration over $\bk$. For the sake of simplicity, we restrict the study of $g_2$ to a one-dimensional geometry where photons are strongly confined also along the $y$ direction. None of the qualitative features is however expected to be modified when going to higher dimensions. In addition to the negative peak at $x=x'$ due to the repulsive interparticle interactions, we see a propagating correlation pattern which is governed by the group velocity $v_{\rm gr} = \partial_k\omega_k$ where $\omega_k$ is given in Eq.~\eqref{disprel}. As in the case of equilibrium condensates~\cite{Carusotto:2009re,Busch:2013sma}, the fast oscillations at large separations are due to the superluminal form of the dispersion relation Eq.~\eqref{disprel} in the high-momentum region. Low-momenta modes $k^2/m < M c^2 $ propagate with a smaller velocity because of the small but finite phonon mass and are responsible for the long-wavelength oscillations that are visible at small $x-x'$. It is worth observing that dissipation introduce an extra dissipative length scale $L_d = c/ \Gamma$ in addition to the usual healing length $\xi =1/mc $: for the parameters of the figures, we have $L_d/\xi_{\rm in} \sim 30$, which means that dissipation affects the profiles of $g_2$ only at large distances.

For the sake of completeness, we conclude the section with a study of the first-order coherence function $g_1$ as defined in Eq.~\eqref{eq:g1def}. In particular, we consider its Fourier-space form 
\begin{equation}
g_{1 ,\bk} (t,t')\! \doteq \int\! d\xx\, \ep{ -i \bk \cdot\xx}g_1 (\xx,t,\xx'= 0,t') ,
\end{equation}
which describes the photon momentum distribution (for $t=t'$) and the photon coherence in momentum space (for generic $t\neq t'$). Experimentally, this quantity can be directly extracted from the far-field angular distribution of the emitted light and its coherence. Using Eq.~\eqref{eq:bogotf}, this quantity is given for $\bk \neq 0 $ by
\begin{equation}
\begin{split}
g_{1 ,\bk} (t,t')=\ep{ i \omega_p (t-t')} \bigg\{   u_k^2 G^{\varphi^\dagger \varphi}(t,t'; \bk) + v_k^2 G^{\varphi^\dagger \varphi}(t',t; \bk) \\
- 2 u_k v_k \Re \left [G^{\varphi \varphi}(t,t'; \bk)\right ] +v_k^2 \ep{- \Gamma\abs{t-t'}} \ep{-i \omega_k (t-t')} \bigg\} .
\end{split}
\end{equation}
When considering the state after a sudden change, the $g_1$ splits analogously $g_2$ in Eq.~\eqref{g2k_split} into its stationary and its decaying parts, 
\begin{equation}
\begin{split}
g_{1 ,\bk} (t,t') \, \ep{ -i \omega_p (t-t')} =& \ep{- \Gamma\abs{t-t'}} \, g_{1 ,\bk}^{\rm st} (t,t') \\
&+ \ep{- \Gamma(t+t')} \, g_{1 ,\bk}^{\rm dec} (t,t').
\end{split}
\end{equation}
Using Eq.~\eqref{eq:diisnb}, the two components define four independent quantities
\begin{subequations}
\begin{align}
g_{1 ,\bk}^{\rm st} &(t,t') = \Re[ B_1 \ep{ -i \omega_{k,+} \abs{t-t'}}] + i B_2 \sin \left [ \omega_{k,+} (t-t')\right ], \\
\begin{split}
g_{1 ,\bk}^{\rm dec} &(t,t') = B_3 (u_{k,+}^2+v_{k,+}^2) \cos\left [ \omega_{k,+}(t-t') \right ]  \\
 &+ \Re [B_4 \ep{ -i \omega_{k,+} (t+t')}]  -i B_3  \sin\left [ \omega_{k,+}(t-t') \right ]. 
 \end{split}
\end{align}
\end{subequations}
given by
\begin{equation}
\begin{split}
B_1 &= u_{k,+}^2 n_{k,\rm f}^b +v_{k,+}^2 (n_{k,\rm f}^b + 1) - 2 u_{k,+} v_{k,+} \bar c_{k,\rm f}^b , \\
B_2 &= n_{k,\rm f}^b-  v_{k,+}^2 , \\
B_3 &= \delta n_{k}^b ,\\
B_4 &= -2 u_{k,+} v_{k,+} c_k^b .
\end{split}
\end{equation}
These encode the six independent real quantities which characterize the correlation functions of Eq.~\eqref{eq:diisnb}. Hence, unlike the $g_{2,\bk}$, the $g_{1 ,\bk}$ fully characterizes the bipartite state $\bk,-\bk$. 

\begin{figure}
\includegraphics[width=1\linewidth]{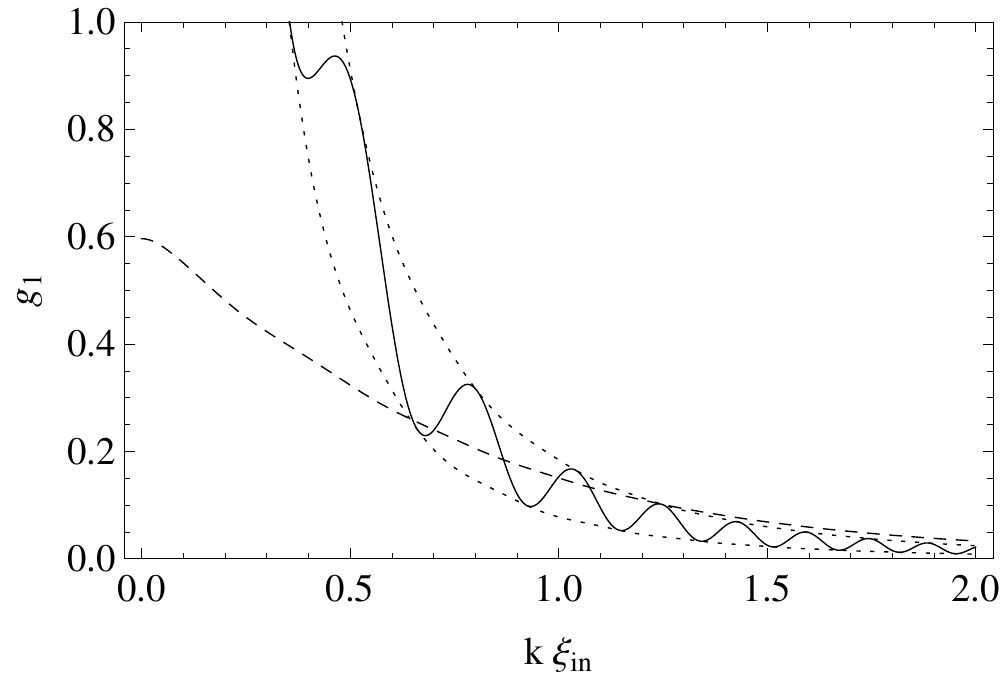}
\caption{Momentum distribution of the cavity photons. The equal-time $g_{1, \bk}(t,t)$ is plotted in momentum space at $t =3/ m c_{\rm in}^2$ (solid line). Dotted lines indicate its lower and upper envelopes. The phonon state is non-separable whenever the lower envelope goes below the dashed line indicating the separability condition \eq{nnsep}. The system and jump parameters are the same as in Fig.~\ref{fig:nkmck} } 
\label{fig:g1ofk}
\end{figure}

In Fig.~\ref{fig:g1ofk}, we represent the equal time $g_{1 ,\bk} (t,t)$, for $t m c_{\rm in}^2= 3 $ and for the same parameters as in the previous figures. Contrary to what was found for $g_{2, \bk}$, the separability threshold of Eq.~\eqref{eq:separability} ($n_k^b = \abs{c_k^b}$) does not simply enter in $g_{1, \bk}$. In fact, to extract it, we need both the upper and lower envelopes of $g_{1 ,\bk} (t,t)$, called respectively $U_k(t)$ and $L_k(t)$. Violation of the inequality 
\begin{equation}
\begin{split}
\label{nnsep}
L_k(t) > \frac{(u_{k,+}-v_{k,+})^2 U_k(t) + 2 v_{k,+}^2}{(u_{k,+}+v_{k,+})^2} 
\end{split}
\end{equation}
implies that the phonon state is nonseparable, i.e., $n_k^b(t) < \abs{c_k^b(t)}$. In the figure, the ratio of \eq{nnsep} is represented by a dashed line. We again see the large domain of $k$ where the phonon state is nonseparable, namely, $k \xi_{\rm in} > 0.6$. 

\section{Conclusions}
\label{sec:Conclusion}

In this article, we studied the quantum fluctuations in coherently pumped and spatially homogeneous photon fluids in planar microcavities. Our attention is focused on the simplest case of a quasiresonant coherent pump at normal incidence on the microcavity, where the photon fluid is at rest and the effective mass of phonon excitations on top of the photon fluid is very small.

When the pump is monochromatic and stationary, the system reaches a stationary state: most remarkably, even if the environment is in its vacuum state, the stationary state of the photon gas is not a vacuum state, but contains a finite occupation of (almost) incoherent phonons. Even though the phonon distribution qualitatively resembles a Planck law at an effective temperature of the order of the interaction energy in the fluid, the nonequilibrium nature of the system leads to quantitatively significant deviations and to violations of the standard fluctuation-dissipation relations.

When the system parameters are suddenly modulated in time, entangled pairs of extra phonons are created in the fluid via processes that are the analog of cosmological pair production or the dynamical Casimir effect. Due to the dissipation, these phonons eventually decay while the system relaxes to a new stationary state. Accurate information on the properties of these extra phonons can be obtained from measurements of the time-dependence of the first- and second-order coherence functions of the cavity photons, which can be used to assess the quantum non-separability of the phonon state after the jump.

The conclusions of this work will provide crucial information in view of studies of the quantum entanglement properties of the Hawking emission of phonons from acoustic black hole horizons in photon fluids.

\acknowledgments
We are grateful to R. Balbinot, D. Gerace, S. Finazzi, and M. Wouters for useful discussions. We also thank S. Robertson and F. Michel for interesting discussions and a careful reading of the manuscript. This work has been supported by ERC through a QGBE Grant and by Provincia Autonoma di Trento. It has also been supported by the French National Research Agency under the Program Investing in the Future Grant No. ANR-11-IDEX-0003-02 associated with the project QEAGE (``Quantum Effects in Analogue Gravity Experiments'').

\appendix

\section{Fluctuation dissipation relation} 
\label{app:FD}
We saw in Sec.~\ref{sec:equilibrium} that the stationary state of phonons when the photon fluid is in its steady state in contact with the environment is not thermal. This might appear at a first glance as quite surprising since under very general conditions, systems weakly interacting with a large stationary reservoir reach a thermal equilibrium state as is guaranteed by the fluctuation-dissipation (FD) relations~\cite{landau1996statistical,Anglin:1992uq}. In this appendix, we shall see that the violation of the FD relation stems from the fact that our system is externally driven by the coherent laser pump with a finite frequency $\omega_p$.

To show that this violation is not due to some approximation, we use the exact Heisenberg equation of motion without performing the Markov approximation used in the body of the text. From Eqs.~\eqref{eq:eomdissip} and~\eqref{eq:GPEgeneral}, in the place of Eq.~\eqref{eq:2x2eom}, the exact equation for linear perturbations is
\begin{equation}
\label{eq:perturb}
\begin{split}
i(\partial_t  &+ \Gamma  )\hat \phi_\bk(t) = \Omega_k \hat \phi_\bk(t)+  m c^2 \hat \phi_{-\bk}^\dagger(t) +
\frac{\hat S_\bk(t)}{\abs{{\mathtt \Phi}_0}} \\
& - i  \int dt'  D(t-t') \ep{i \omega_p(t-t')} \left ( \hat \phi_\bk(t') - \hat \phi_\bk(t) \right ) .
\end{split}
\end{equation}
Using Eq.~\eqref{eq:Sdef} to express $\hat S_\bk$ in terms of the destruction operators $\hat c(\bk,\zeta)$ of the environment, and working in Fourier transform to exploit the stationarity of the situation, the equation takes the form
\begin{equation}
\label{eq:omegaeom}
\begin{split}
O_1 (\omega)\,  \hat \phi_\bk^\omega &+ O_2(\omega) \left (\hat \phi_{-\bk}^{-\omega}\right )^\dagger \\
&=  \int  \! d\zeta g_\zeta \delta({\omega + \omega_p - \omega_\zeta})  \hat c(\bk,\zeta) .
\end{split}
\end{equation}
Using the complex conjugated equation for $-\omega,-\bk $ to eliminate $(\phi_{-\bk}^{-\omega})^\dagger$, we get 
\begin{equation}
\label{eq:solphiomega}
\begin{split}
&\bigg( O_1 (\omega)\,  O_1^*(-\omega) -O_2(\omega) \, O_2^*(-\omega) \bigg) \hat \phi_\bk^\omega =  \\
&\quad \int d\zeta  g_\zeta 
\bigg( \delta({\omega +\omega_p - \omega_\zeta}) O_1^*(-\omega) \hat c(\bk,\zeta) \\
&\quad \hspace{1cm} + \delta({\omega  -  \omega_p + \omega_\zeta}) O_2(\omega) \hat c(-\bk,\zeta)^\dagger \bigg ) .
\end{split}
\end{equation}
Making the Bogoliubov transformation of \eq{eq:bogotf} to get the equation for the phonon field $\hat \varphi_\bk^\omega$ simply amounts to replacing in the above equation $O_i$ by $u_k O_i - v_k O_{3-i}$, for $i \in \{1,2\}$. Hence, the same type of expression applies to $\hat \varphi_\bk^\omega$, or, more generally, to any linear superposition (even $\omega$ dependent) of $\hat \phi_\bk^\omega $ and $(\hat \phi_{-\bk}^{-\omega})^\dagger $.

We now remind the reader that the FD relation trivially applies at the level of the operators of the environment. Namely, when working in a thermal state, one has
\begin{equation}
\begin{split}
\frac{{\rm Tr}\left (\hat \rho  \left \{ \hat c(\bk,\zeta), \hat c(\bk,\zeta)^\dagger \right \} \right )}{\left [ \hat c(\bk,\zeta), \hat c(\bk,\zeta)^\dagger \right ]} =  \coth \frac{\beta \omega_\zeta }{2}, 
\label{FDenv}
\end{split}
\end{equation}
as can be immediately verified by computing the commutator and the expectation value of the anticommutator of $\hat c(\bk,\zeta)$ and $\hat c(\bk',\zeta')^\dagger $.

When the pump frequency $\omega_p = 0$, the situation is simple: Because of the Dirac $\delta$ function in \eq{eq:solphiomega}, and because the energy of the environment modes $\omega_\zeta $ is positive for all $\zeta$, for $\omega>0$, $\hat \phi_\bk^\omega$ is driven only by the destruction operator $\hat c(\bk,\zeta)$ with $\omega_\zeta = \omega$. Then, using \eq{FDenv}, a direct evaluation gives 
\begin{equation}
\begin{split}
\frac{{\rm Tr}\left (\hat \rho \, \left \{ \hat \phi_\bk^\omega,(\hat \phi_\bk^{\omega})^\dagger \right \} \right )}{\left [ \hat \phi_\bk^\omega,(\hat \phi_\bk^{\omega})^\dagger \right ]} = \coth \frac{\beta \omega}{2} ,
\end{split}
\end{equation}
irrespective of the values of $O_1(\omega)$ {\it and} $O_2(\omega)$. This is the standard FD relation. 

When $\omega_p \neq 0$, to get a concise expression, as in Ref.~\cite{Caldeira:1981rx}, it is useful to introduce the effective density of states $J(\omega)$ through $d\zeta\, g_\zeta^2 = d\omega_\zeta \,J(\omega_\zeta)$. A direct evaluation then gives 
\begin{widetext}
\begin{equation}
\label{eq:FDviolated}
\begin{split}
\frac{\rm{Tr}\left(\hat \rho \,  \{ \hat \phi_\bk^\omega,  \hat \phi_\bk^{\dagger,\omega} \} \right)}{[ \hat \phi_\bk^\omega,  \hat \phi_\bk^{\dagger,\omega} ]}  &= \frac{ \abs{ O_1(-\omega) }^2 J(\omega_p + \omega) \coth\beta(\omega_p+\omega)/2 -  \abs{O_2 (\omega)}^2  J(\omega_p - \omega) \coth\beta(\omega_p - \omega)/2 }{\abs{O_1(-\omega)}^2 J(\omega_p + \omega) - \abs{O_2 (\omega)}^2  J(\omega_p - \omega)} .\\
\end{split}
\end{equation}
\end{widetext}
We see that a FD relation is recovered only if $O_2(\omega) J(\omega_p - \omega) = 0$. In such a case, the argument in the $\coth$ in the right hand side of \eq{eq:FDviolated} is displaced as if there were a chemical potential $\mu = - \omega_p$.

For a general environment, all frequencies $\omega_\zeta$ are positive and cover the whole $\omega>0$ region, so $J(\omega)$ vanishes only for $\omega<0$. As a result, the $O_2(\omega) J(\omega_p - \omega) = 0$ condition requires either working at very high frequencies $\omega > \omega_p$ outside the region of interest for quantum hydrodynamics, or having $O_2(\omega) = 0$, that is a vanishing interaction between photons.

\section{Separability and Cauchy Schwartz inequalities}
\label{app:CS}

The notion of nonseparability for a two-mode system has been introduced by Werner~\cite{Werner:1989}. A state is defined as separable when it can be written as a statistical superposition of products of two one-mode states. For homogeneous systems, the two modes correspond to the $\pm\bk$ components of some field, and the density matrix $\rho_k$ in the $k$-th two-mode subspace is separable if it can be written as
\begin{equation}
\begin{split}
\rho_{k} &= \sum_n p_n \rho_{n,\bk}\otimes \rho_{n,-\bk},
\end{split}
\end{equation}
where $p_n > 0$ are probabilities, and $\rho_{n,i}$ are the density matrices of quantum states for the $\pm \bk$ subsystems. Because the Bogoliubov transformation of Eq.~\eqref{eq:bogotf} mixes the $\bk$ and $-\bk $ sectors, it may happen that a state is separable if viewed in terms of photon operators $\hat a_\bk$ but nonseparable if viewed in terms of phonon operators $\hat \varphi_\bk$, and viceversa. As an example, the $\Gamma \to 0$ stationary state of Sec.~\ref{sec:equilibrium} is indeed separable in term of the phonon operators $\hat \varphi_\bk$, and non-separable in term of the photon operators $\hat a_\bk$.

We now show that the nonseparability criterion of Eq.~\eqref{eq:separability} based on the phonon $\hat \varphi_\bk$ operators is equivalent to the violation of a Cauchy-Schwarz (CS) inequality for phonon operators. We consider the modified equal-time second-order correlation which is obtained from the standard photonic one $g_{2,\bk}(t,t')$ by
\begin{equation}
\begin{split}
g_{2,\bk}^b(t,t') &\doteq g_{2,\bk}(t,t')+2 v_k(u_k-v_k)\Re [\hat\varphi_\bk(t) , \hat \varphi_\bk^\dagger(t')]\\
&=  2 \left (u_k-v_k  \right )^2 \Re[G(t,t', \bk)];
\end{split} 
\end{equation} 
subtraction of the contribution of the commutator in this expression is equivalent to taking the normal ordering with respect to the phonon operators $\hat b_\bk$ of \eq{eq:phidec}, hence the $b$ superscript in the above notation. In terms of this quantity, the CS inequality reads 
\begin{equation}
\label{eq:CS}
\begin{split}
\mathcal{D}_k(t,t')=\frac{g_{2 ,\bk}^b (t,t) g_{2 ,\bk}^b (t',t')-|{g_{2 ,\bk}^b (t,t')}|^2}{4 \left (u_k-v_k  \right )^4} \geq 0 .
\end{split}
\end{equation}
No violation of Eq.~\eqref{eq:CS} can occur in classical statistical physics. In the absence of dissipation, the phonon mean occupation number $n_k^b$ and correlation term $c_k^b$ are both well defined, and constant before and after a sudden jump. Using these two quantities, one obtains
\begin{equation}
\label{eq:CSdifferece}
\begin{split}
\mathcal{D}_\bk(t,t') = \left [  (n_k^b)^2 - \abs{c_k^b}^2\right ] \sin^2[\omega_k (t-t')]. 
\end{split}
\end{equation}
Hence, the CS inequality is violated if and only if the state is non separable (when $\sin[\omega_k (t-t')]\neq 0$).

\begin{figure}
\includegraphics[width =\linewidth]{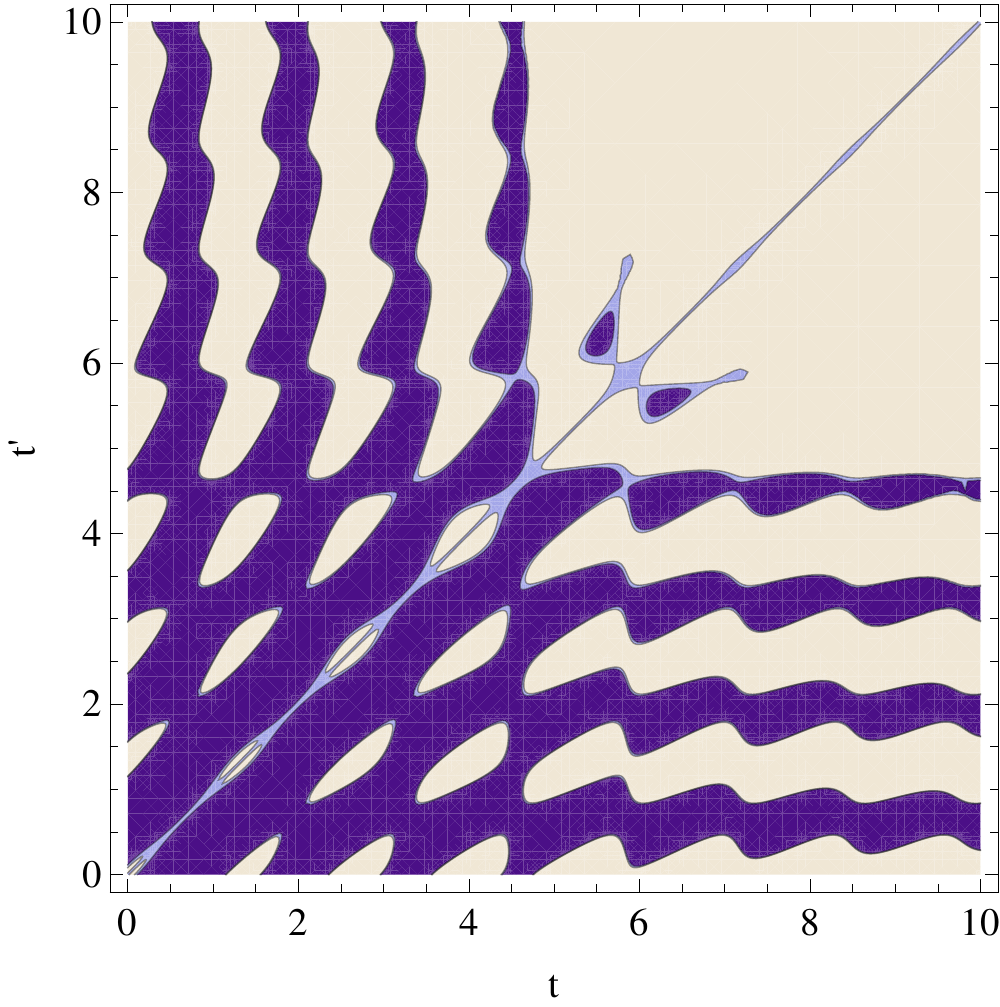}
\caption{Plot of the Cauchy-Schwarz criterion Eq.~\eqref{eq:CS} as a function of $t,t'$ for $k= 1.5 m c_{\rm in}$. Dark blue regions indicate values below $-10^{-4}$ that significantly violate the inequality. White regions indicate values larger than $10^{-4}$. Light blue regions indicate values close to $0$. The system and jump parameters are the same as in Fig.~\ref{fig:nkmck}.}
\label{fig:2DCS}
\end{figure} 

In the presence of dissipation, as discussed in the main text of the article and in Sec.~IV~B of Ref.~\cite{Busch:2013sma}, the coupling of the phonon field $\hat\varphi$ to an environment introduces intrinsic ambiguities in the definition of nonseparability. Nevertheless, decomposition $\hat\varphi$ at any time $t$ over {\it instantaneous} destruction and creation operators $\hat b_\bk, \hat b^\dagger_{-\bk}$ allows one to show that for $\Gamma(t - t') \ll 1$, the above relation between the sign of $\mathcal{D}_\bk(t,t')$ and the nonseparability criterion based on these operators remains valid to leading order in $\Gamma/\omega_k$. Accepting this inherent uncertainty of order $\Gamma/\omega_k \ll 1$, one can then follow how nonseparability is progressively lost as time goes on. This physics is illustrated in Fig.~\ref{fig:2DCS}: the quantity in Eq.~\eqref{eq:CS} displays three different behaviors depending on the values of $t,t'$ compared to the characteristic time $t_{\rm loss}$ of \eq{tloss}. 
\begin{enumerate}
\item
For $(t,t')\gg t_{\rm loss}$, no violation is observed, because the state is separable, as expected.

\item
For $t \ll  t_{\rm loss} \ll t' $, \eq{eq:CS} can be violated only for $t$ such that $g_{2,\bk}^b(t,t)<0$, because $g_{2,\bk}^b(t',t')>0$. Along a constant-$t'$ cut, the reader will recognize the behavior already seen in Fig.~\ref{fig:g2oft}.

\item
For $t,t' \ll t_{\rm loss}$ and $ \Gamma/ \omega_k^2 \ll  |t-t'| \lesssim  1/ \omega_k $, \eq{eq:CS} is violated. This is the most robust regime for non separability.
\end{enumerate}
Note that besides the transition from point $2$ to point $3$, there is always a narrow band $ |t-t'| \ll \Gamma/ \omega_k^2$ where the inequality is never violated with a positiveness of order $\Gamma / \omega_k$. As a result, a two-time measurement of $g_{2,\bk}^b(t\neq t')$ is required to identify non-separable states. 

We conclude this appendix with a short discussion of the standard momentum-space CS inequality for photon $\hat a_\bk$ operators; see~\cite{deNova:2012hm} or~\cite{PhysRevLett.108.260401} for its atomic counterpart. In terms of the momentum-space second-order photon coherence\footnote{
Note that this quantity should not be confused with $g_{2,\bk}(t,t')$ defined in Eq.~\eqref{g2k} as the Fourier transform of the real-space $g_2(\xx,t,\xx',t')$.}
\begin{equation}
\begin{split}
 \mathcal{G}_2(\bk,\bk') = { \rm Tr} \left (\hat \rho  a^\dagger_\bk a^\dagger_{\bk'} a_\bk a_{\bk'}\right ).
\end{split}
\end{equation}
Its explicit form is
\begin{equation}
\label{eq:CSfora}
\begin{split}
[\mathcal{G}_2(\bk, \bk')]^2 \leq \mathcal{G}_2(\bk,\bk)\, \mathcal{G}_2(\bk',\bk') , 
\end{split}
\end{equation}
Physically, this quantity describes the correlations between the fluctuations of the photon occupation numbers in the modes $\bk$ and $\bk'$. Thanks to the Gaussian nature of the state, we can apply the Wick theorem and expand this expression in terms of quadratic operators. For homogeneous states, we get 
\begin{equation}
 \begin{split}
 \mathcal{G}_2(\bk,\bk')  &=  \delta_{\bk+\bk'}\abs{c_k^a}^2 +  \delta_{\bk-\bk'} (n_k^a)^2+ n^a_k n^a_{k'}.
 \end{split}
 \end{equation} 
For $\bk' = - \bk$, the CS condition Eq.~\eqref{eq:CSfora} is then equivalent to the separability condition Eq.~\eqref{eq:separability} applied to photon operators. Inserting the explicit form of the photon correlations \eq{photCS}, it is immediate to see that the CS inequality for the photon field is indeed violated  for $\bk'=-\bk$ even in the stationary state.

Note that this result is not peculiar to the driven-dissipative case, but is also found in the ground state of equilibrium Bogoliubov theory. It has a straightforward physical interpretation if one recalls that the finite-$\bk$ photons originate from a parametric scattering process where two pump photons at $\bk_p=0$ scatter into the $\pm\bk$ states.

\bibliographystyle{../../biblio/h-physrev}
\bibliography{../../biblio/bibliopubli} 

\end{document}